\documentclass[aps,prd,letterpaper,superscriptaddress,showpacs,amsmath,amssymb]{revtex4}

\usepackage[dvips]{pstcol}
\usepackage{fancybox}
\usepackage{fancyhdr}
\usepackage{epsfig}
\usepackage{setspace}
\usepackage{dcolumn}

\newcommand{\gtsim}{\raisebox{-1ex}{\mbox{$\mbox{}\stackrel{\textstyle >}{\sim}\mbox{}$}}}

\begin{document}

\begin{flushright} 
FERMILAB-PUB-09-069-E
\end{flushright}

\title{Measurement of the $b$-hadron Production Cross
Section Using Decays to $\mu^- D^0 X$ Final States 
in $p\bar{p}$ Collisions
at $\sqrt{s}=1.96\, \rm TeV$}


\affiliation{Institute of Physics, Academia Sinica, Taipei, Taiwan 11529, Republic of China} 
\affiliation{Argonne National Laboratory, Argonne, Illinois 60439} 
\affiliation{University of Athens, 157 71 Athens, Greece} 
\affiliation{Institut de Fisica d'Altes Energies, Universitat Autonoma de Barcelona, E-08193, Bellaterra (Barcelona), Spain} 
\affiliation{Baylor University, Waco, Texas  76798} 
\affiliation{Istituto Nazionale di Fisica Nucleare Bologna, $^x$University of Bologna, I-40127 Bologna, Italy} 
\affiliation{Brandeis University, Waltham, Massachusetts 02254} 
\affiliation{University of California, Davis, Davis, California  95616} 
\affiliation{University of California, Los Angeles, Los Angeles, California  90024} 
\affiliation{University of California, San Diego, La Jolla, California  92093} 
\affiliation{University of California, Santa Barbara, Santa Barbara, California 93106} 
\affiliation{Instituto de Fisica de Cantabria, CSIC-University of Cantabria, 39005 Santander, Spain} 
\affiliation{Carnegie Mellon University, Pittsburgh, PA  15213} 
\affiliation{Enrico Fermi Institute, University of Chicago, Chicago, Illinois 60637}
\affiliation{Comenius University, 842 48 Bratislava, Slovakia; Institute of Experimental Physics, 040 01 Kosice, Slovakia} 
\affiliation{Joint Institute for Nuclear Research, RU-141980 Dubna, Russia} 
\affiliation{Duke University, Durham, North Carolina  27708} 
\affiliation{Fermi National Accelerator Laboratory, Batavia, Illinois 60510} 
\affiliation{University of Florida, Gainesville, Florida  32611} 
\affiliation{Laboratori Nazionali di Frascati, Istituto Nazionale di Fisica Nucleare, I-00044 Frascati, Italy} 
\affiliation{University of Geneva, CH-1211 Geneva 4, Switzerland} 
\affiliation{Glasgow University, Glasgow G12 8QQ, United Kingdom} 
\affiliation{Harvard University, Cambridge, Massachusetts 02138} 
\affiliation{Division of High Energy Physics, Department of Physics, University of Helsinki and Helsinki Institute of Physics, FIN-00014, Helsinki, Finland} 
\affiliation{University of Illinois, Urbana, Illinois 61801} 
\affiliation{The Johns Hopkins University, Baltimore, Maryland 21218} 
\affiliation{Institut f\"{u}r Experimentelle Kernphysik, Universit\"{a}t Karlsruhe, 76128 Karlsruhe, Germany} 
\affiliation{Center for High Energy Physics: Kyungpook National University, Daegu 702-701, Korea; Seoul National University, Seoul 151-742, Korea; Sungkyunkwan University, Suwon 440-746, Korea; Korea Institute of Science and Technology Information, Daejeon, 305-806, Korea; Chonnam National University, Gwangju, 500-757, Korea} 
\affiliation{Ernest Orlando Lawrence Berkeley National Laboratory, Berkeley, California 94720} 
\affiliation{University of Liverpool, Liverpool L69 7ZE, United Kingdom} 
\affiliation{University College London, London WC1E 6BT, United Kingdom} 
\affiliation{Centro de Investigaciones Energeticas Medioambientales y Tecnologicas, E-28040 Madrid, Spain} 
\affiliation{Massachusetts Institute of Technology, Cambridge, Massachusetts  02139} 
\affiliation{Institute of Particle Physics: McGill University, Montr\'{e}al, Qu\'{e}bec, Canada H3A~2T8; Simon Fraser University, Burnaby, British Columbia, Canada V5A~1S6; University of Toronto, Toronto, Ontario, Canada M5S~1A7; and TRIUMF, Vancouver, British Columbia, Canada V6T~2A3} 
\affiliation{University of Michigan, Ann Arbor, Michigan 48109} 
\affiliation{Michigan State University, East Lansing, Michigan  48824}
\affiliation{Institution for Theoretical and Experimental Physics, ITEP, Moscow 117259, Russia} 
\affiliation{University of New Mexico, Albuquerque, New Mexico 87131} 
\affiliation{Northwestern University, Evanston, Illinois  60208} 
\affiliation{The Ohio State University, Columbus, Ohio  43210} 
\affiliation{Okayama University, Okayama 700-8530, Japan} 
\affiliation{Osaka City University, Osaka 588, Japan} 
\affiliation{University of Oxford, Oxford OX1 3RH, United Kingdom} 
\affiliation{Istituto Nazionale di Fisica Nucleare, Sezione di Padova-Trento, $^y$University of Padova, I-35131 Padova, Italy} 
\affiliation{LPNHE, Universite Pierre et Marie Curie/IN2P3-CNRS, UMR7585, Paris, F-75252 France} 
\affiliation{University of Pennsylvania, Philadelphia, Pennsylvania 19104}
\affiliation{Istituto Nazionale di Fisica Nucleare Pisa, $^z$University of Pisa, $^{aa}$University of Siena and $^{bb}$Scuola Normale Superiore, I-56127 Pisa, Italy} 
\affiliation{University of Pittsburgh, Pittsburgh, Pennsylvania 15260} 
\affiliation{Purdue University, West Lafayette, Indiana 47907} 
\affiliation{University of Rochester, Rochester, New York 14627} 
\affiliation{The Rockefeller University, New York, New York 10021} 
\affiliation{Istituto Nazionale di Fisica Nucleare, Sezione di Roma 1, $^{cc}$Sapienza Universit\`{a} di Roma, I-00185 Roma, Italy} 

\affiliation{Rutgers University, Piscataway, New Jersey 08855} 
\affiliation{Texas A\&M University, College Station, Texas 77843} 
\affiliation{Istituto Nazionale di Fisica Nucleare Trieste/Udine, I-34100 Trieste, $^{dd}$University of Trieste/Udine, I-33100 Udine, Italy} 
\affiliation{University of Tsukuba, Tsukuba, Ibaraki 305, Japan} 
\affiliation{Tufts University, Medford, Massachusetts 02155} 
\affiliation{Waseda University, Tokyo 169, Japan} 
\affiliation{Wayne State University, Detroit, Michigan  48201} 
\affiliation{University of Wisconsin, Madison, Wisconsin 53706} 
\affiliation{Yale University, New Haven, Connecticut 06520} 
\author{T.~Aaltonen}
\affiliation{Division of High Energy Physics, Department of Physics, University of Helsinki and Helsinki Institute of Physics, FIN-00014, Helsinki, Finland}
\author{J.~Adelman}
\affiliation{Enrico Fermi Institute, University of Chicago, Chicago, Illinois 60637}
\author{T.~Akimoto}
\affiliation{University of Tsukuba, Tsukuba, Ibaraki 305, Japan}
\author{B.~\'{A}lvarez~Gonz\'{a}lez$^s$}
\affiliation{Instituto de Fisica de Cantabria, CSIC-University of Cantabria, 39005 Santander, Spain}
\author{S.~Amerio$^y$}
\affiliation{Istituto Nazionale di Fisica Nucleare, Sezione di Padova-Trento, $^y$University of Padova, I-35131 Padova, Italy} 

\author{D.~Amidei}
\affiliation{University of Michigan, Ann Arbor, Michigan 48109}
\author{A.~Anastassov}
\affiliation{Northwestern University, Evanston, Illinois  60208}
\author{A.~Annovi}
\affiliation{Laboratori Nazionali di Frascati, Istituto Nazionale di Fisica Nucleare, I-00044 Frascati, Italy}
\author{J.~Antos}
\affiliation{Comenius University, 842 48 Bratislava, Slovakia; Institute of Experimental Physics, 040 01 Kosice, Slovakia}
\author{G.~Apollinari}
\affiliation{Fermi National Accelerator Laboratory, Batavia, Illinois 60510}
% ...

\collaboration{CDF Collaboration\footnote{With visitors from $^a$University of Massachusetts Amherst, Amherst, Massachusetts 01003,
$^b$Universiteit Antwerpen, B-2610 Antwerp, Belgium, 
$^c$University of Bristol, Bristol BS8 1TL, United Kingdom,
$^d$Chinese Academy of Sciences, Beijing 100864, China, 
$^e$Istituto Nazionale di Fisica Nucleare, Sezione di Cagliari, 09042 Monserrato (Cagliari), Italy,
$^f$University of California Irvine, Irvine, CA  92697, 
$^g$University of California Santa Cruz, Santa Cruz, CA  95064, 
$^h$Cornell University, Ithaca, NY  14853, 
$^i$University of Cyprus, Nicosia CY-1678, Cyprus, 
$^j$University College Dublin, Dublin 4, Ireland,
$^k$University of Edinburgh, Edinburgh EH9 3JZ, United Kingdom, 
$^l$University of Fukui, Fukui City, Fukui Prefecture, Japan 910-0017
$^m$Kinki University, Higashi-Osaka City, Japan 577-8502
$^n$Universidad Iberoamericana, Mexico D.F., Mexico,
$^o$Queen Mary, University of London, London, E1 4NS, England,
$^p$University of Manchester, Manchester M13 9PL, England, 
$^q$Nagasaki Institute of Applied Science, Nagasaki, Japan, 
$^r$University of Notre Dame, Notre Dame, IN 46556,
$^s$University de Oviedo, E-33007 Oviedo, Spain, 
$^t$Texas Tech University, Lubbock, TX  79609, 
$^u$IFIC(CSIC-Universitat de Valencia), 46071 Valencia, Spain,
$^v$University of Virginia, Charlottesville, VA  22904,
$^w$Bergische Universit\"at Wuppertal, 42097 Wuppertal, Germany,
$^{ee}$On leave from J.~Stefan Institute, Ljubljana, Slovenia, 
}}
\noaffiliation



\begin{abstract}
 We report  a measurement of the production cross section
for $b$ hadrons in $p\bar{p}$ collisions at $\sqrt{s}=1.96\, \rm TeV$.
 Using a data sample derived from an integrated luminosity 
 of $83\, \rm pb^{-1}$ 
 collected with the upgraded Collider Detector (CDF\,II) 
at the Fermilab Tevatron,
 we analyze $b$ hadrons, $H_b$, partially reconstructed
in the semileptonic decay 
mode $H_b \rightarrow \mu^-  D^0 X$.
Our measurement of the 
inclusive production cross section for $b$ hadrons with
transverse momentum $p_T > 9 \, {\rm GeV}/c$ and 
rapidity $|y|<0.6$ is  
$\sigma = 1.30\, \mu{\rm b} \pm 0.05\, \mu {\rm b} {\rm (stat)} \pm 0.14 \, 
\mu {\rm b} {\rm (syst)} \pm 0.07 \mu {\rm b} ({\mathcal{B})}$, where
the uncertainties are statistical, systematic, and from branching
fractions, respectively.  The differential cross sections
$d\sigma/dp_T$ are found to be in good agreement with
recent measurements of the $H_b$ cross section and 
well described by
fixed-order next-to-leading logarithm predictions.

\end{abstract}



\maketitle

\section{Introduction}

Measurements of the $b$-hadron  production cross section in 
high energy $p\bar{p}$  collisions 
provide an excellent test of perturbative quantum
chromodynamics (QCD).
Next-to-leading order (NLO) calculations of the 
$b$-hadron cross section \cite{NDE1, NDE2} have been available 
for more than a decade.
These NLO calculations have 
been recently
supplemented with corrections for terms proportional 
to $\alpha_s^2 [\alpha_s \log (p_T/m_b )]^k$ 
and $\alpha_s^3 [\alpha_s \log (p_T/m_b )]^k$ with $k \geq 1$, 
where $\alpha_s$ is the strong coupling constant, $p_T$ is
the transverse momentum of the $b$ hadron, and $m_b$ is the
mass of the bottom quark.  These calculations are  
referred to as fixed order next-to-leading logarithm (FONLL) 
\cite{FONLL}.

A number of $b$-hadron cross-section measurements has been performed
at the Tevatron.   In the 1992-1996 running period, Tevatron Run~I,
measurements were performed at $\sqrt{s}= 1.8\, \rm TeV$ 
by the CDF and D0 experiments 
using several different techniques and  final 
states \cite{run1, semilep, Abe1993, AbachiMu1995}.  
These measurements were consistently higher 
	than theoretical expectations from NLO calculations.

Since 2001, the Tevatron has been running at $\sqrt{s}= 1.96\, \rm TeV$.   
To date, the CDF Collaboration has made two inclusive 
$b$-hadron cross-section measurements
at the higher center-of-mass energy. 
The first was an inclusive measurement using 
$H_b \to J/\psi X, J/\psi \to \mu^+ \mu^-$ decays, where $H_b$ 
denotes 
a $b$ hadron \cite{AcostaJpsi2005}.   This measurement was
performed for $|y|<0.6$ \cite{rapidity} and was the first 
to map out the $b$-hadron cross section 
down to zero transverse momentum. 
The $H_b\to J/\psi X$ cross-section  
result, along with an improved FONLL calculation \cite{FONLL},
prompted studies \cite{FONLL2, others2} suggesting that prior discrepancies
had been resolved with improved measurement and calculational
techniques.

More recently, the CDF Collaboration 
has performed a measurement of the $B^+$ meson
cross section using $740\, \rm pb^{-1}$ of data through the
fully reconstructed decay chain
$B^+ \to J/\psi K^+$, with $J/\psi \to \mu^+ \mu^-$, for $B^+$ transverse
momentum, $p_T(B^+)$, greater than $6\, \rm GeV/{\it c}$ and 
$|y|<1$.   With 
large statistics and a high purity signal, the total
uncertainty on this measurement is smaller than $10\%$ \cite{run2jpsik}.  The result 
is also in good agreement with FONLL calculations.  The difference between 
the original NLO calculation and the current FONLL calculation
arises from a number of different factors, as discussed in 
Ref.~\cite{FONLL2}.

While improved theoretical calculations
compare favorably to recent measurements, 
more experimental input is needed.  
In particular, 
cross-section 
measurements using different $H_b$ decay modes can be complementary 
to one another since they might be acquired using different trigger 
paths and be subject to different background contributions.
For example,
Ref. \cite{ThatPaoloPaper} showed that previous $b$-hadron cross-section
measurements were inconsistent when comparing semileptonic and
$J/\psi$ final states.

In this paper, we present 
the first measurement of the $H_b$ cross section using
semileptonic decays at $\sqrt{s}=1.96\, \rm TeV$.    This 
analysis takes advantage of the distinct semileptonic $b$-hadron
decay signature and provides a measurement that is complementary
to those made in $H_b\rightarrow J/\psi X$ modes.
We reconstruct signals of semileptonic $b$-hadron decays,
$H_b \to \mu^- D X$,  where charm mesons, $D^0$ and 
${D^*}^+$, are
reconstructed in fully hadronic modes, using a data 
sample acquired with a dedicated semileptonic trigger path.
Throughout this paper, 
any reference to a specific charge state implies the antiparticle
state as well.

Previous measurements of the $b$-hadron 
cross section at the Tevatron using semileptonic decays were 
performed using inclusive 
$H_b \to \mu^- X$ \cite{Abe1993,AbachiMu1995} and 
the semielectronic $H_b\to e^- D^0 X$ \cite{semilep} modes.  There
are no previously published $b$-hadron cross-section 
results from the Tevatron
utilizing the 
$H_b \to \mu^- D^0 X$ mode.  We improve upon prior
semileptonic results with a 
comparable data sample by extending the measurement to 
lower values of $p_T(H_b)$ than have been accessible 
in previous analyses.  The trigger path used in this analysis, which will
be described in detail below, accepts muons with $p_T > 4\, {\rm GeV}/c$ while
previous measurements were limited to leptons with $p_T > 8 \, {\rm GeV}/c$.

We begin in  Sec.~\ref{sec:overview} with an overview of
the measurement technique.  The CDF~II detector and trigger system will
be described in Sec.~\ref{sec:detector}, followed by a discussion
of trigger selection in Sec.~\ref{sec:trigger}.  
Section~\ref{DataSampleSection} will describe the event selection
and candidate reconstruction.
Sections~\ref{sec:eff} and \ref{sec:acc} then describe the measurements of 
signal reconstruction efficiency  and acceptance, followed by an assessment 
of the background 
contributions in Sec.~\ref{sec:bkg}.   
In Sec.~\ref{sec:results}, we present
our results along with a comparison to theoretical 
predictions.

\section{Overview of the Measurement}
\label{sec:overview}

We select events with an identified
muon in conjunction with a fully reconstructed charm 
meson ($D^0$ or ${D^*}^+$, generically referred to as 
a $D$ meson) such that the $\mu^- D$ topology is consistent
with a $b$-hadron decay.  To
measure the cross section, we count the number of signal
events and correct for detector acceptance, 
identification inefficiency,
and contributions from background sources.  
Incorporating these factors, along with the measured 
$p\bar{p}$ luminosity, ${\cal L}$, we extract the
cross section, $\sigma$,  
times branching ratio ($\mathcal{B}$) in the 
following way:

\begin{equation}
\sigma (p \overline{p} \to H_b X) 
\times \mathcal{B} = \frac{N (1-f_{bck})}
{ 2 \alpha \epsilon \mathcal{L}} \ \  , \label{CrossSectionEquation} 
\end{equation}
where $N$ is the number of observed candidate  $\mu^- D^0$ ($\mu^- {D^*}^+$) 
events and $f_{bck}$ is the background fraction, \it i.e. \rm the
fraction of events 
not originating from the decay of a $b$ hadron.  
The acceptance $\alpha$ is defined as the
fraction of events in which the muon and the charm decay
particles pass through active regions of the CDF II detector.
The detection efficiency $\epsilon$ 
is the fraction of these events that are reconstructed.

The product of
branching ratios, $\mathcal{B}$, is shorthand for
$\mathcal{B}(H_b\rightarrow \mu^- D^0 X)\times 
\mathcal{B}(D^0\rightarrow K^-\pi^+)$
in the $\mu^- D^0$ mode and 
$\mathcal{B}(H_b\rightarrow \mu^- {D^*}^+ X)\times 
\mathcal{B}({D^*}^+\rightarrow
D^0 \pi^+)\times \mathcal{B}(D^0\rightarrow K^-\pi^+)$ for the 
$\mu^- {D^*}^+$ mode.  
The generic $b$ hadron, $H_b$, is an admixture 
of all weakly decaying $b$ hadrons, including $B^+$, $B^0$,
$B^0_s$, $B^+_c$, and $\Lambda_b$.  To extract the $H_b$
cross section, we utilize the measurement of the inclusive
$H_b \to \mu^- D^0 X$ branching ratio from the DELPHI and OPAL
experiments \cite{delphi}.  To apply this 
branching ratio to Tevatron data, we assume that the 
fragmentation fractions at high energy ($p_T(H_b)>9\, {\rm GeV}/c$)
are the same at LEP 
and the Tevatron.  Charm branching ratios are taken from 
CLEO, ARGUS, and ALEPH measurements \cite{D0Br, DsBr}.

Under the assumption that $b$ and $\overline{b}$ quarks are 
produced in equal rates at the Tevatron, we 
include both $\mu^+ \overline{D}^0$ 
and $\mu^- D^{0}$ final  states and introduce the factor
of 2 in the denominator of Eq.~(\ref{CrossSectionEquation})
to report
the cross section for a single $b$-hadron flavor.

To extract the number of candidate signal events, $N$, 
we reconstruct  $H_b \to \mu^-  D^{0} X$ with $D^0 \to K^- \pi^+$,  
as well as  
$H_b \to \mu^-  {D^*}^+ X$ with  ${D^*}^+ \to D^0 \pi^+$  and 
$D^0 \to K^- \pi^+$,
events using data that were acquired 
via a dedicated semimuonic $H_b$ trigger
that requires a well measured muon 
and a track displaced from the primary vertex.    
 The ${D^*}^+$ sample is a subset of the 
$D^0\to K^- \pi^+$ sample.   We use these two samples to make
separate, but correlated, measurements.

The selection criteria are chosen to 
preferentially select $H_b$ decays and
reject combinatoric backgrounds.
There are two types of physics backgrounds that
exhibit a $\mu^- D$ signature that is similar
to that of the expected signal.
One arises from  direct
$p\bar{p}\rightarrow c \overline{c}X$ 
production where one charm quark decays
to a muon and the other fragments into a 
 $D$ meson.  The
second physics background arises from $p\bar{p}\rightarrow
b \overline{b}X$ 
events where one $b$ hadron decays to a muon while the 
other decays to a $D$ meson.  The contributions from 
these sources are assessed using
a combination of data and Monte Carlo (MC) simulation.

The acceptance  $\alpha$  is determined by using 
MC simulation.  We generate $b$ quarks based 
on input distributions in transverse momentum $p_T$ and 
rapidity $y$   
taken from theoretical calculations 
as well as previous cross-section 
measurements.  Fragmentation and decay are performed as part of the 
MC simulation.
The passage of generated decay 
	products in the CDF II detector is then simulated.  Based upon
the output 
of these simulated events, the fraction of
produced events that fall within the active detector volume is
calculated.

We factorize the determination of the trigger and detector 
efficiency, $\epsilon$, into a product of relative 
efficiencies and an 
absolute efficiency.
The advantage of this technique is that we can measure 
relative efficiencies from the data (as will
be described in Sect.~\ref{sec:eff}) and minimize our
reliance on simulation.

The sensitivity of this analysis is limited by the systematic 
uncertainties.   To limit our dependence on 
theoretical input and on detector simulation we  
estimate systematic uncertainties from data
wherever possible.

\section{The CDF~II Detector}
\label{sec:detector}

The CDF~II detector comprises a 
charged-particle tracker in the magnetic field of a
solenoid surrounded by a full coverage calorimeter 
and by muon detectors. In this section we give a 
short description of those detector components 
that are relevant to this analysis. A more detailed 
description of them and of the entire detector 
is given in 
Refs.~\cite{CDF, Sill1, Affolder1, Ascoli1, CLC}. 

We utilize a cylindrical coordinate system with the $z$ axis
oriented in the direction of the proton beam and the 
$r$-$\phi$ plane perpendicular to the beamline.    We also
define a polar angle, $\theta$, measured from the $z$ axis 
and pseudorapidity
is defined as $\eta =-\ln[\tan(\theta/2)]$.

Two devices inside the 1.4 T solenoidal magnetic field
 are used for measuring the 
momentum of charged 
particles: the silicon vertex detector (SVX~II) 
and the central tracking chamber (COT). 
The SVX~II consists of double-sided microstrip 
sensors arranged in five cylindrical shells 
with radii between $2.5$ and $10.6\, \rm cm$. The SVX~II detector 
is divided into three contiguous 
sections along the beam direction for a total $z$ 
coverage of $90\, \rm  cm$. The COT is a cylindrical 
drift chamber containing 96 sense wire layers 
grouped into eight alternating superlayers 
of axial and stereo wires. The active volume 
covers $|z|< 155\, \rm cm$ and $43$ to $132\, \rm  cm$ in 
radius. 

The central muon detector (CMU) surrounds 
the central electromagnetic and hadronic calorimeters.
The calorimeters have a depth corresponding to  5.5 
interaction lengths.
The CMU detector covers a pseudorapidity 
range $|\eta|<$ 0.6  and is segmented 
into two barrels of 24 modules, each module 
covering $12.6^\circ$ 
in $\phi$.  Every module is further segmented into three 
submodules, each covering $4.2^\circ$  in $\phi$ and 
consisting of four layers of drift chambers. The smallest 
drift unit, called a stack, covers 
$1.2^\circ$ in $\phi$.  Adjacent pairs of stacks 
are combined together into a tower.  A track 
segment is identified by at least two hits 
out of the four layers of a stack. 
A second set of muon drift chambers, 
the CMP, covers the same $\eta$-range as the CMU and 
is located behind an additional 
steel absorber of $\sim\! 3.3$ interaction lengths. Muons 
that produce a track segment in both the CMU and 
CMP systems are called CMUP muons.  A third set of 
muon drift chambers, the CMX, covers a 
pseudorapidity range $0.6<|\eta|<1.0$ and is separated by
6.2 interaction lengths from the nominal interaction 
point.

Luminosity is measured using gas
Cherenkov counters (CLC) mounted at
small angles to the beamline.  The CLC measures 
the rate of inelastic $p \overline{p}$ collisions. 
The inelastic $p \overline{p}$ cross section at 
$\sqrt{s} = 1960$ GeV is scaled from measurements 
at $\sqrt{s} =$ 1800 GeV using the calculations 
in Ref. \cite{CLCref}. The integrated 
luminosity is determined with negligible statistical
uncertainty and a 6\% systematic accuracy \cite{LumiErr}.

The CDF experiment uses a three-level trigger system. At 
level~1 (L1), data from every beam crossing are 
stored in a pipeline capable of buffering data 
from 14 beam crossings. The level~1 trigger either 
rejects events or copies them onto one of the 
four level~2 (L2) buffers. Events that pass 
the level~1 and level~2 selection criteria are sent to the 
level~3 (L3) trigger, a cluster of computers 
running a speed-optimized version of the offline
event reconstruction code.

To select heavy flavor events, we rely heavily upon
charged-particle tracking in the trigger.  At level~1, charged
tracks with $p_T\ge 1.5\, \rm GeV/{\it c}$ and 
$|\eta|<1$ are found by the extremely fast 
tracker (XFT) \cite{xft},
which uses information from the axial wires in the
COT to perform $r$-$\phi$ track finding with high
efficiency ($>90\%$) and good transverse momentum 
resolution ($\delta p_T/p_T^2 = 0.017$).
The track extrapolation
system (XTRP) takes the track information from the
XFT and provides extrapolation 
information so that the XFT tracks can
be matched to track segments found in the muon 
detectors \cite{xtrp}.  Tracks 
are matched to track segments in the CMU, CMP and CMX
to identify muon candidates at level~1.

At level~2, the Silicon Vertex Trigger (SVT) combines information
from the axial layers of the SVX~II with the charged-particle 
tracking
information from the XFT.   Owing to the high precision 
information from the silicon detector, the SVT provides a 
precise measurement of the track
impact parameter $d_0$ 
(distance of closest approach to the 
beamline)
in the transverse plane 
with a
resolution of approximately
$50\, \rm \mu m$.  This resolution arises from 
approximately equal
contributions of  $35\, \rm \mu m$ from  
resolution and intrinsic spread of the position of the  
$p\overline{p}$ interaction. 
Tracks with large impact parameter are 
utilized to identify heavy flavor decay and strongly
reject light ($u,d,s$) quark events.

For the data sample utilized in this analysis,
level~3 event reconstruction included full COT tracking 
but did not include tracking in the silicon detectors.
Events accepted by the level~3 trigger are stored for 
subsequent analysis.

The flexibility of the CDF~II trigger permits the selection of
samples with no leptons (all hadronic modes), one lepton 
(semileptonic modes) and two leptons ($J/\psi\rightarrow 
\mu^+\mu^-$ and $\Upsilon\rightarrow \mu^+\mu^-$
events.)  For this analysis we utilize all three types of
triggers, and we describe them in the following section.

\section{Trigger Paths}
\label{sec:trigger}

A single trigger path is defined as a specific set
of level~1, level~2, and level~3 selection criteria 
in the CDF~II trigger system. 
The primary trigger path utilized in this analysis 
is referred to as the 
$\mu$SVT path.   Data acquired through other trigger paths 
are utilized
in this analysis to determine the selection
efficiencies described in Sect.~\ref{sec:eff}.

\bf {\boldmath $\mu$}SVT Path. \rm 
  At level~1, 
this path identifies a muon candidate with 
$p_T > 4 \, {\rm GeV}/c$ by requiring
that an XFT track be matched to segments 
in the CMU and CMP detectors.
At level~2 the trigger utilizes information from the
SVT to identify displaced tracks.  We require a single
track (which is not the muon candidate) with $p_T > 2 \, {\rm GeV}/c$ and
an impact parameter between 120 $\mu \, \rm m$ and $1\, \rm mm$ to 
be identified, where 
the impact parameter measured by the SVT is relative to the 
beamline
as determined over a large sampling of events.
At level~3, full event reconstruction is done, except 
for track reconstruction in  SVX~II.  In order for the 
event to be accepted, 
level~3 requires confirmation of the muon identified at
level~1 and confirmation of the XFT track 
identified at level~2.
The 
invariant mass of the two tracks must be less than 
$5\, {\rm GeV}/c^2$, the azimuthal opening angle between the
two tracks, $\Delta \phi$,  
must be $2^\circ < \Delta \phi<$ 90$^\circ$, 
and the tracks must be consistent with coming from the same
$p\bar{p}$ interaction vertex.

\bf Inclusive Muon Path. \rm This trigger requires that one
muon candidate with $p_T > 8\, {\rm GeV}/c$ 
be identified by the XFT, CMU, and CMP.  This path provides
a sample of $J/\psi \rightarrow \mu^+\mu^-$ decays where
only one of the muons is required in the trigger.  The 
second muon in the event is unbiased, and can be utilized
to directly measure the efficiency of the trigger selection.

\bf Heavy Flavor Path. \rm  This trigger path is 
for hadronic decays of $H_b$ and $D$
hadrons.  Two oppositely charged 
XFT tracks with $p_T>2\, \rm GeV/{\it c}$ and 
$\Delta \phi<135^\circ$ are required at level~1.  Those 
tracks are then required to have $120 \, \mu{\rm m}<d_0< 1\, {\rm mm}$
as measured by the SVT at level~2.  The level~3 selection
requires confirmation of the level~2 quantities using 
full reconstruction.  From this path, we
select large 
samples of $D^+ \rightarrow K^- \pi^+ \pi^+$, 
$D^+ \rightarrow \phi \pi^+ \rightarrow K^+ K^- \pi^+$, and
$D^+_s \rightarrow \phi \pi^+ \rightarrow K^+ K^- \pi^+$ decays.  Since
only two of the three tracks are required in the trigger,
the third track is unbiased with respect to the trigger.  We can
therefore use this sample to measure the efficiency
of the SVT.

\bf $H_b$ Semileptonic Backup Path. \rm This trigger path is used to 
determine the level~3 selection efficiency.  
Its requirements are the same as the $\mu$SVT trigger, except that it 
has no requirements at level~3, and only a fraction of 
events that pass the level~1 and level~2 trigger requirements
are recorded.

\bf $J/\psi$ Dimuon Path. \rm The $J/\psi$ trigger paths which require one
muon in the CMU system and a second muon in either the CMU or CMX system
are used to measure 
the SVX~II offline efficiency.  The $J/\psi$ CMU-CMU 
trigger requires, at level~1, 
at least two CMU muons be found with $p_T > 1.5\, {\rm GeV}/c$.  There are 
no additional selection criteria applied  
at level~2.  At level~3, an opposite sign muon pair is required 
with invariant mass 
between $2.7 \, {\rm GeV}/c^2$ and 
$4.0\, {\rm GeV}/c^2$, which is efficient for $J/\psi$ and 
$\psi(2s)$ decays into dimuons.  The two muon tracks 
must also have 
$|\Delta z_0| <5 \, \rm cm$ and $\Delta \phi_0 <$ 130$^\circ$, where
$\Delta z_0$ is the separation of the two 
tracks along the beamline ($r=0$) and $\Delta \phi_0$ is the
separation of the two tracks in azimuth at $r=0$.  The $J/\psi$ 
CMU-CMX trigger 
is similar, except that instead of two CMU muons, one CMU and 
one CMX muon 
must be identified.  The
CMX muon must have $p_T > 2.0\, {\rm GeV}/c$.

\section{Event Selection \label{DataSampleSection}}

This analysis uses data 
derived from $83\, \rm pb^{-1}$ of $p\overline{p}$
collisions at $\sqrt{s}=1.96\, \rm  TeV$ 
that were collected between October 2002 and May 2003.  Thanks 
to the trigger utilized in this analysis, this 
relatively small sample of data is sufficient to significantly 
improve upon previous cross-section measurements in 
semileptonic channels, and yields a result that is limited 
by the systematic uncertainties.  

We begin the signal reconstruction by requiring the
candidate muon, pion, and kaon tracks  
to satisfy COT fiducial and quality selection criteria, which 
include requirements on the number of hits assigned to the track
and a loose selection on fit quality of the track.  
Muon selection criteria are motivated by the trigger
requirements.  The muon track must have $p_T>4\, {\rm GeV}/c$ and  
$|\eta|<0.6$ and must be matched to CMU and 
CMP track segments that satisfy fiducial and quality selection criteria, 
which include requirements on the number of hits used in the track 
segment and the quality of the match between the central track and the 
CMP track segment \cite{thesis}.  
The candidate muon must also be matched to a CMUP  muon 
found by the level~1 trigger.  

In order to reduce combinatoric background, the pion and the kaon 
candidates must have  $p_T>1.0\, {\rm GeV}/c$ and  
$|\eta|<1.0$,
must satisfy SVX~II fiducial and quality cuts, which include 
requirements on the number of hits assigned to the track and
a loose selection on the fit quality of the track \cite{thesis}. 
The two tracks 
must both come from a common displaced vertex and be 
consistent 
with the decay of a charmed meson.
In addition, at least one of the candidate tracks 
must be matched to an SVT track 
that passes the displaced-track-trigger requirements, 
have $120 \, \mu{\rm m}<d_0< 850 \, \mu{\rm m}$, 
and have hits in all of the SVX~II layers used 
by the SVT.  The pion and kaon candidates as well as the 
pion and muon candidates must have 
opposite charges.  We combine the pion and kaon 
candidates to form $D^0\to K^-\pi^+$
candidates.  We form the combined momentum vector
of the muon and the $D^0$ candidate and require that this vector
have $|\eta(\mu^- D^0)|<0.6$.
The number of $D^0$ signal events is determined by fitting the 
$K^- \pi^+$ invariant mass distribution 
between $1.74 \, {\rm GeV}/c^2$ and $1.98\, {\rm  GeV}/c^2$ to 
a Gaussian peak plus a linear background.

\begin{figure}
\psfig{figure=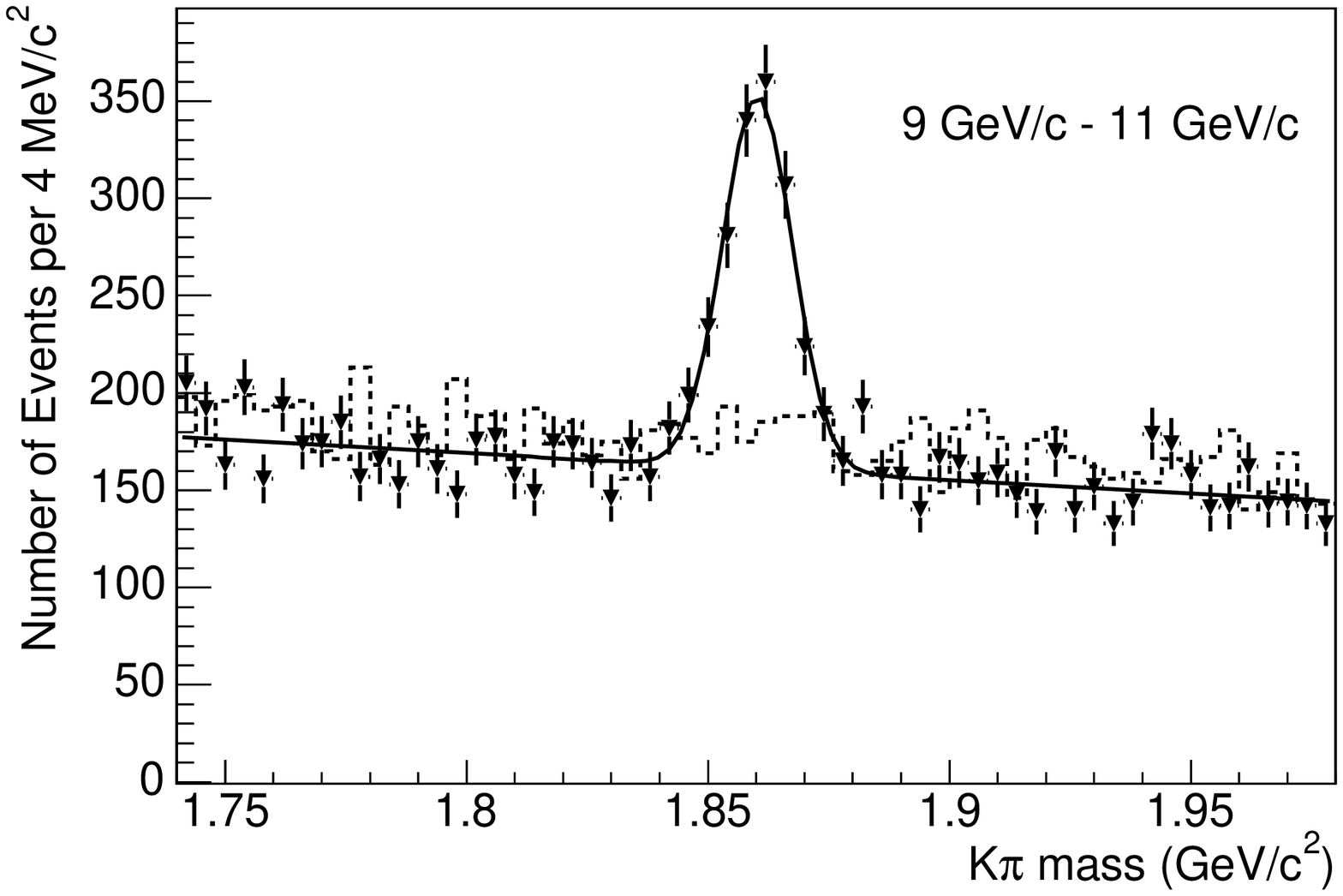,width=0.48\textwidth}
\psfig{figure=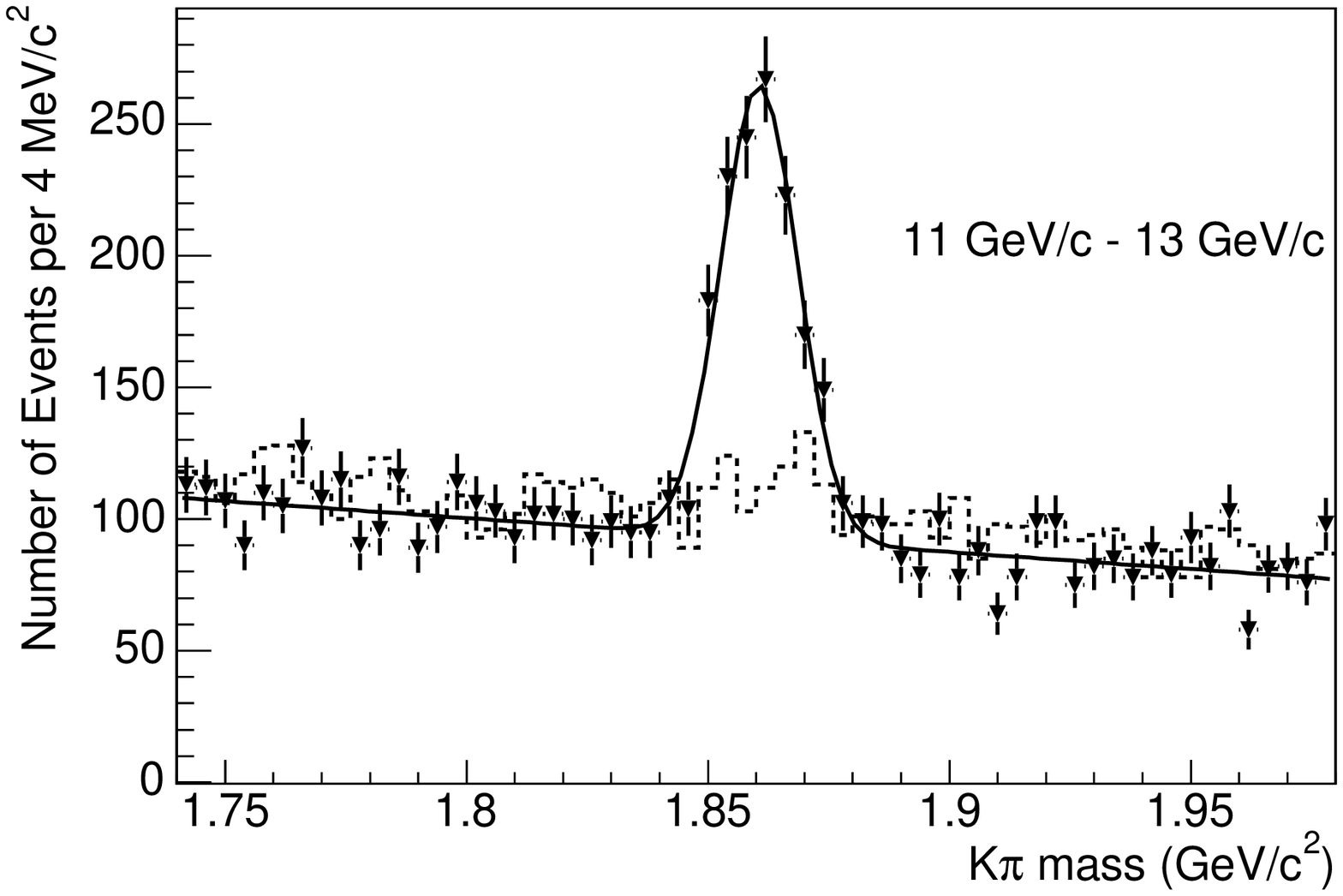,width=0.48\textwidth}
\psfig{figure=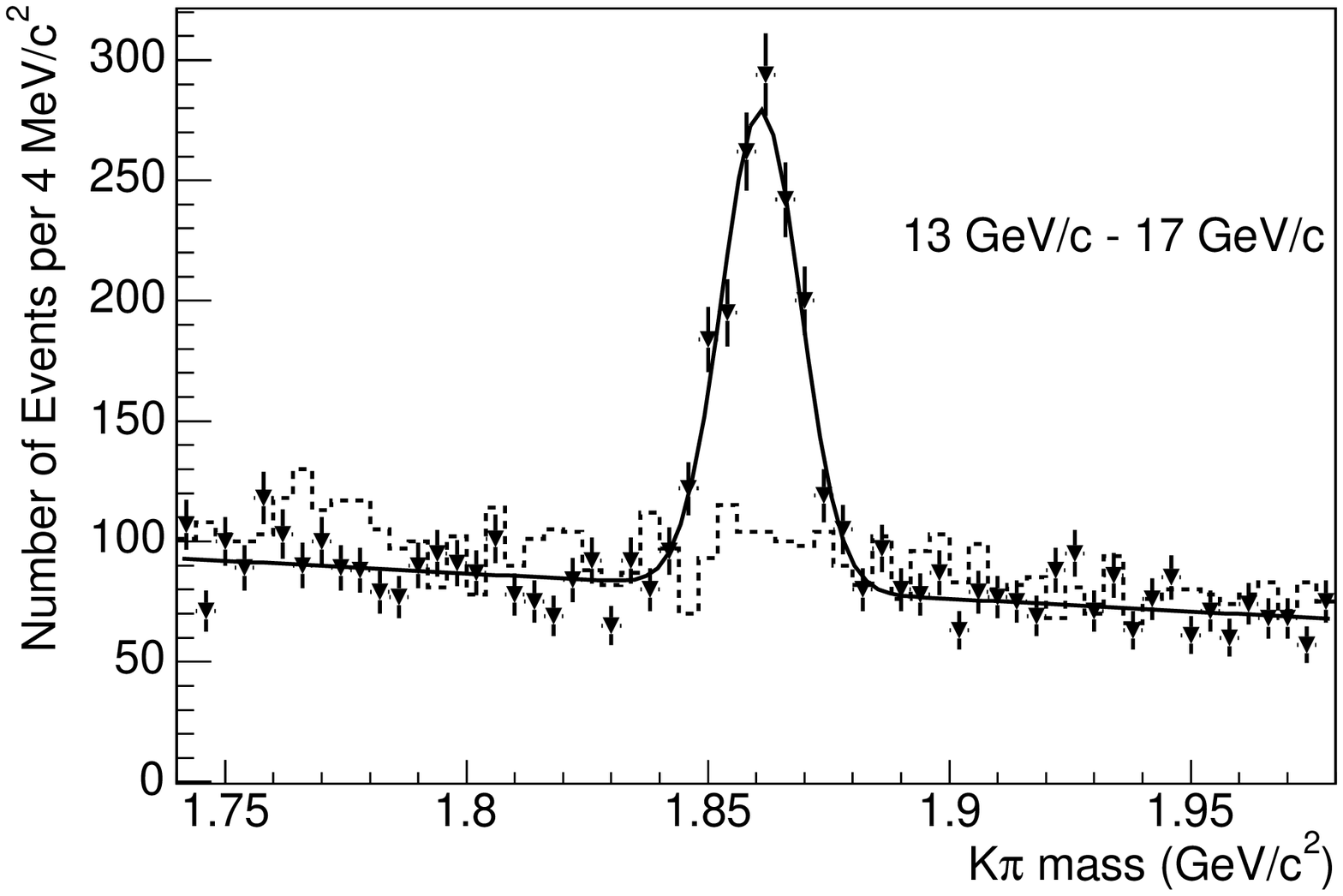,width=0.48\textwidth}
\psfig{figure=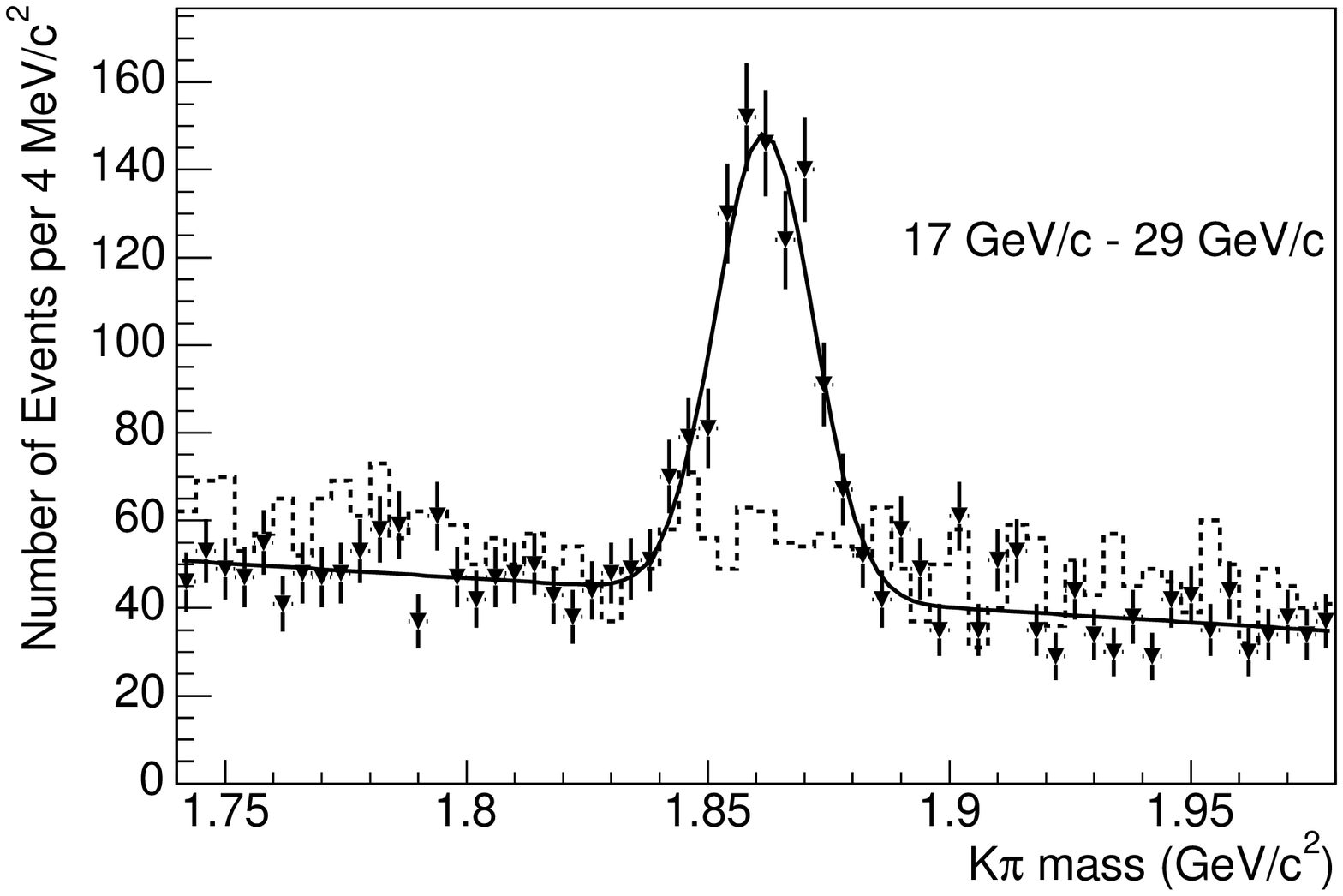,width=0.48\textwidth}
\psfig{figure=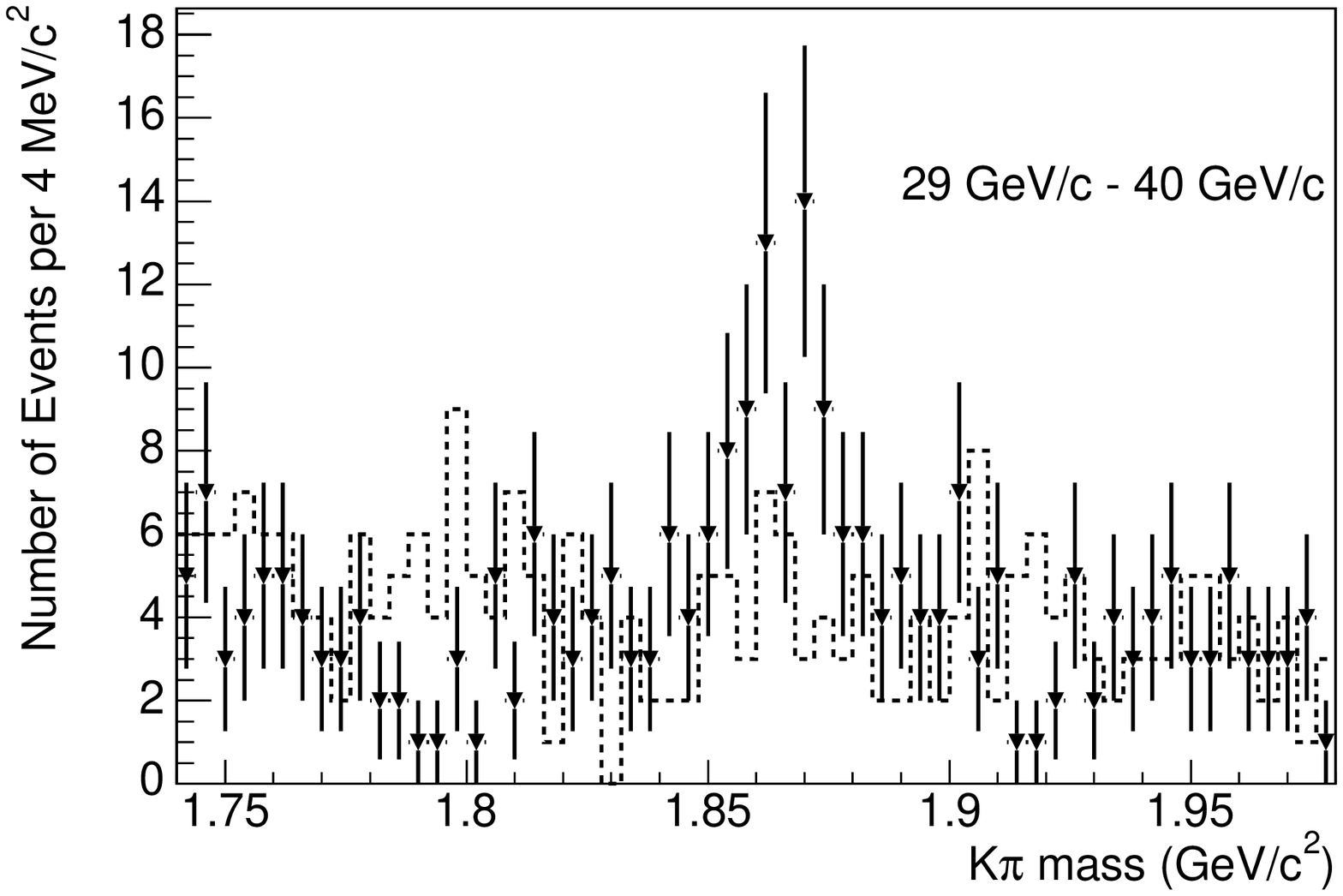,width=0.48\textwidth}
\psfig{figure=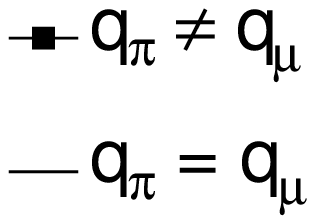,width=0.48\textwidth}
\caption{The $\pi^+ K^-$ mass for different ranges of $p_T(\mu^- K^- \pi^+)$.
Points indicate events where the muon and pion have opposite charges, 
which is the right-sign correlation, and the dashed histograms 
contain events where the 
muon and pion have the 
same charge, which is the wrong-sign correlation.  The solid line is 
a fit to the right-sign data using a Gaussian plus linear background.
The $\chi^2$ per degree of freedom for these fits range from 0.9 to 1.2.  
Because of low statistics, the yield in the highest $p_T$ bin is measured by 
sideband subtraction. \label{D0fitmassplot}}
\end{figure}

\begin{figure}
\psfig{figure=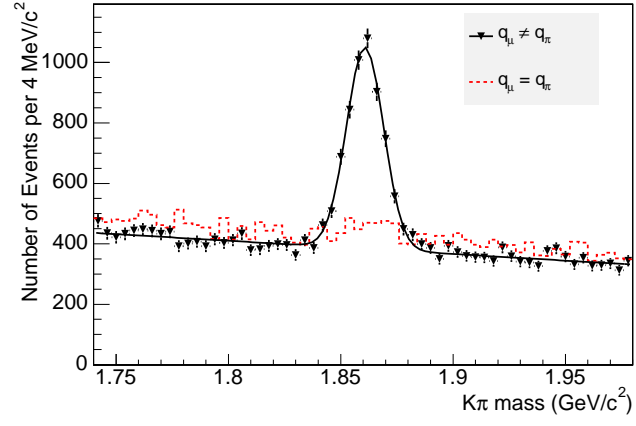,width=0.50\textwidth}
\caption{The $\pi^+ K^-$ mass for $p_T(\mu^- K^-\pi^+)>$ 9.0 GeV/$c$.  The solid histogram shows events with the right-sign correlation between muon and 
pion, and the dashed histogram shows events where the muon and pion have the wrong-sign correlation.  The solid line is 
a fit to the right-sign data using a Gaussian plus linear background.
\label{D0fitmassplotTotalNoSVT}}
\end{figure}

The $D^0$ mass plots 
are shown for different regions of $p_T(\mu^- D^0)$ in 
Fig.~\ref{D0fitmassplot}.  While the signal yield is rather 
low
in the highest $p_T$ region, clean signals are observed in 
the lower
$p_T$ regions.
The signal yields are reported in Table~\ref{yieldtable}.  
Figure~\ref{D0fitmassplotTotalNoSVT} 
shows the $D^0$ mass distribution
for all events with $p_T(\mu^- D^{0})>$ 9.0 GeV/$c$ and
$|\eta(\mu^- D^0)|<0.6$.

\begin{figure}
\psfig{figure=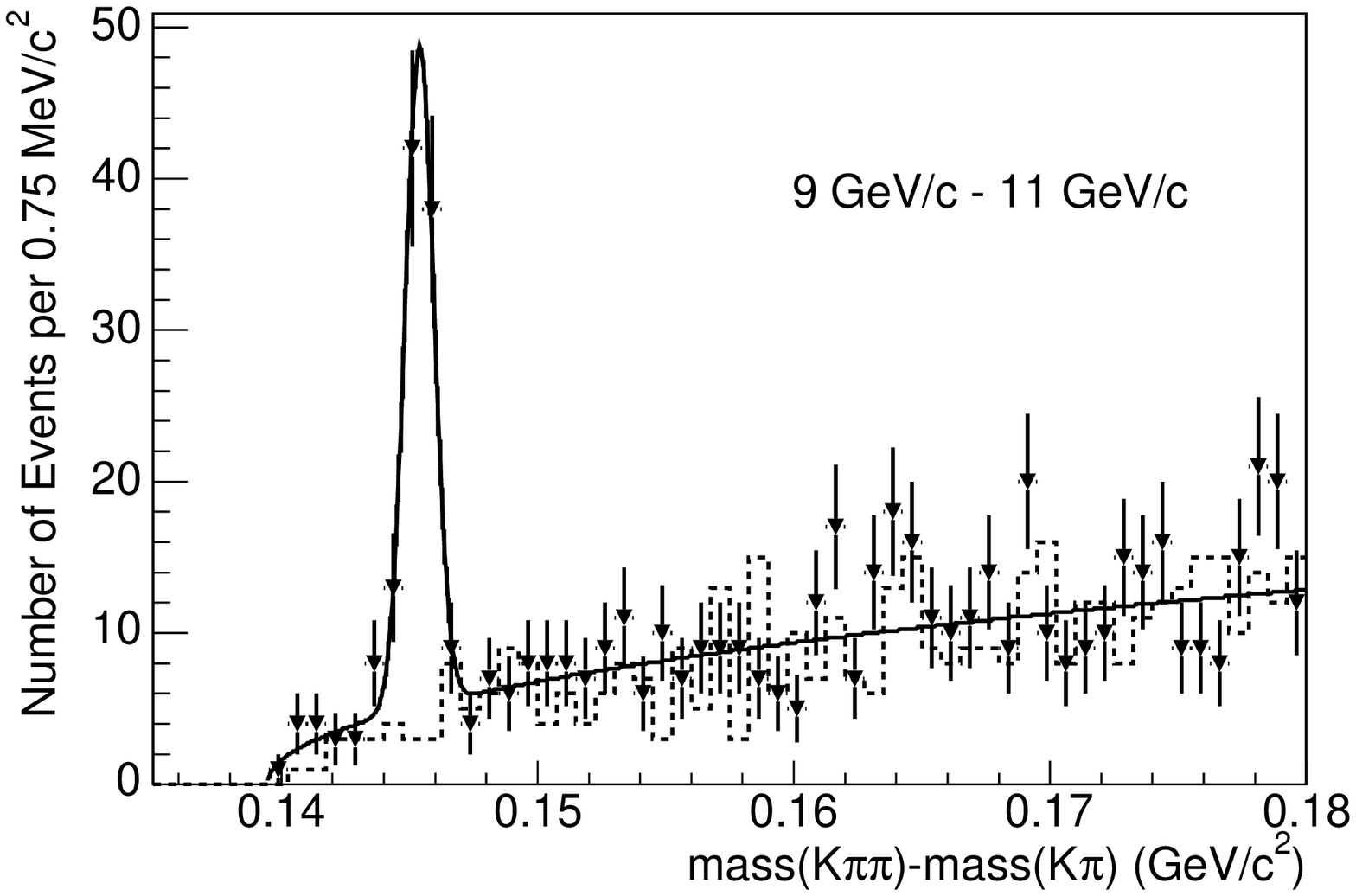,width=0.48\textwidth}
\psfig{figure=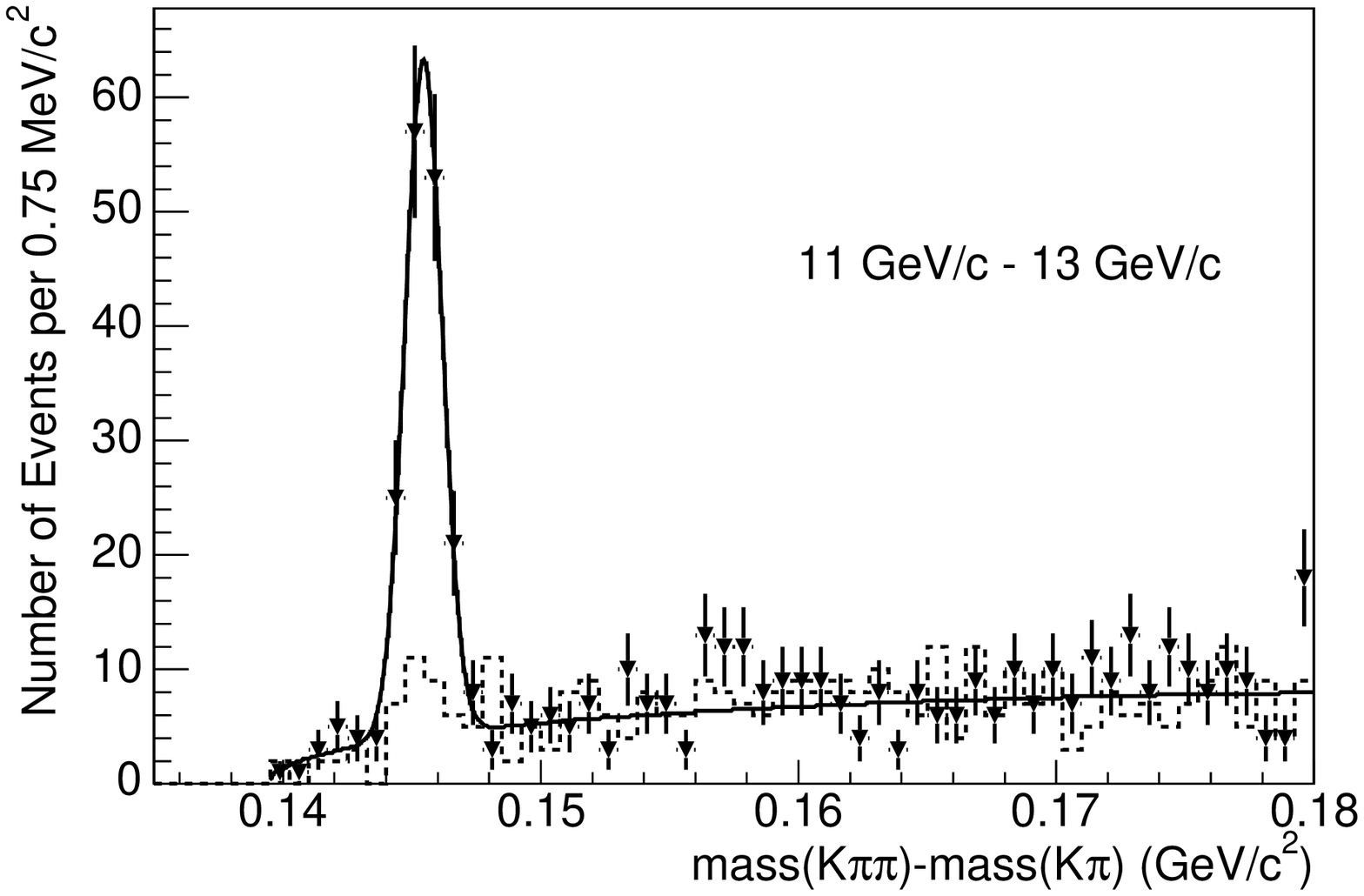,width=0.48\textwidth}
\psfig{figure=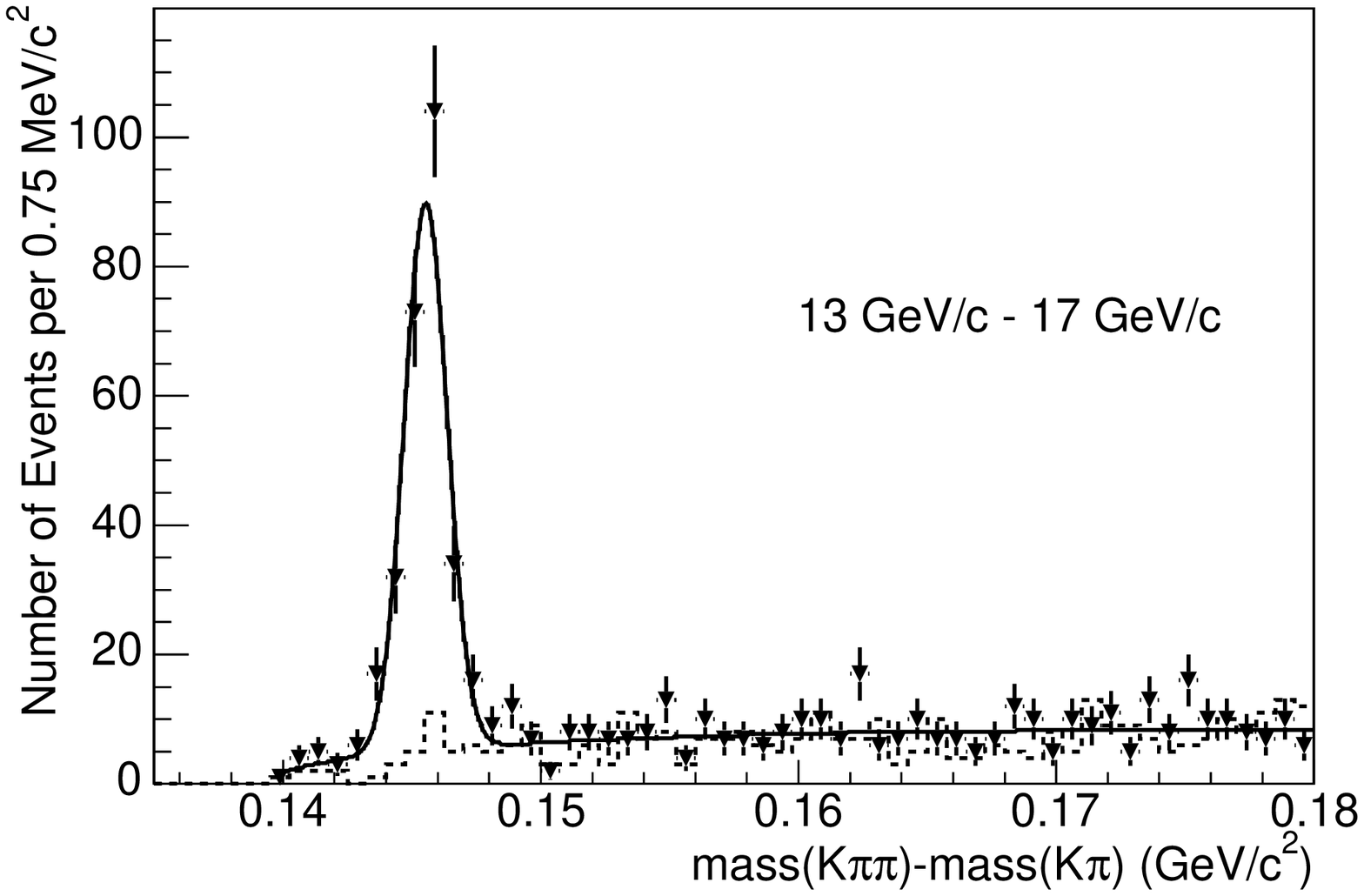,width=0.48\textwidth}
\psfig{figure=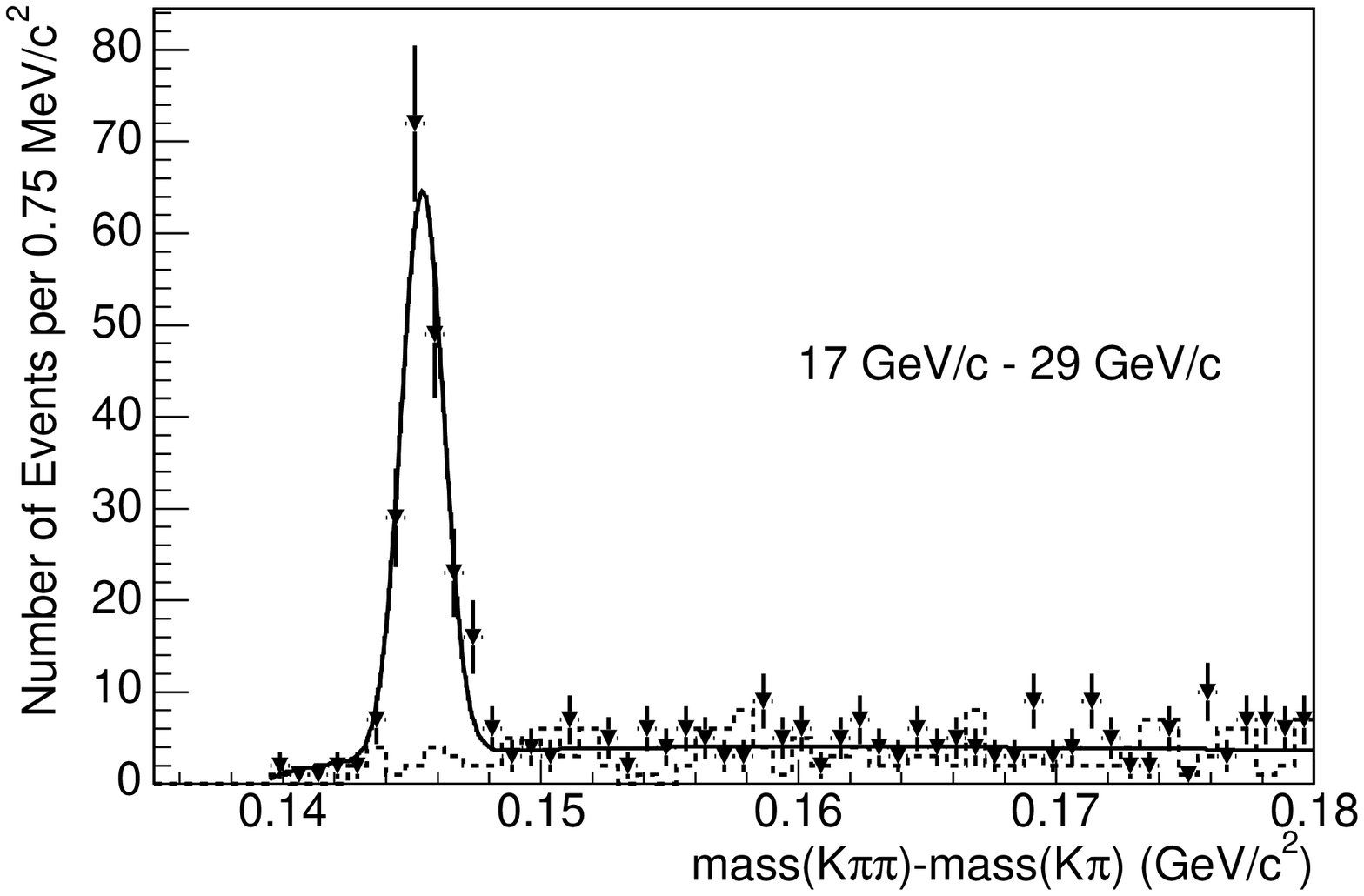,width=0.48\textwidth}
\psfig{figure=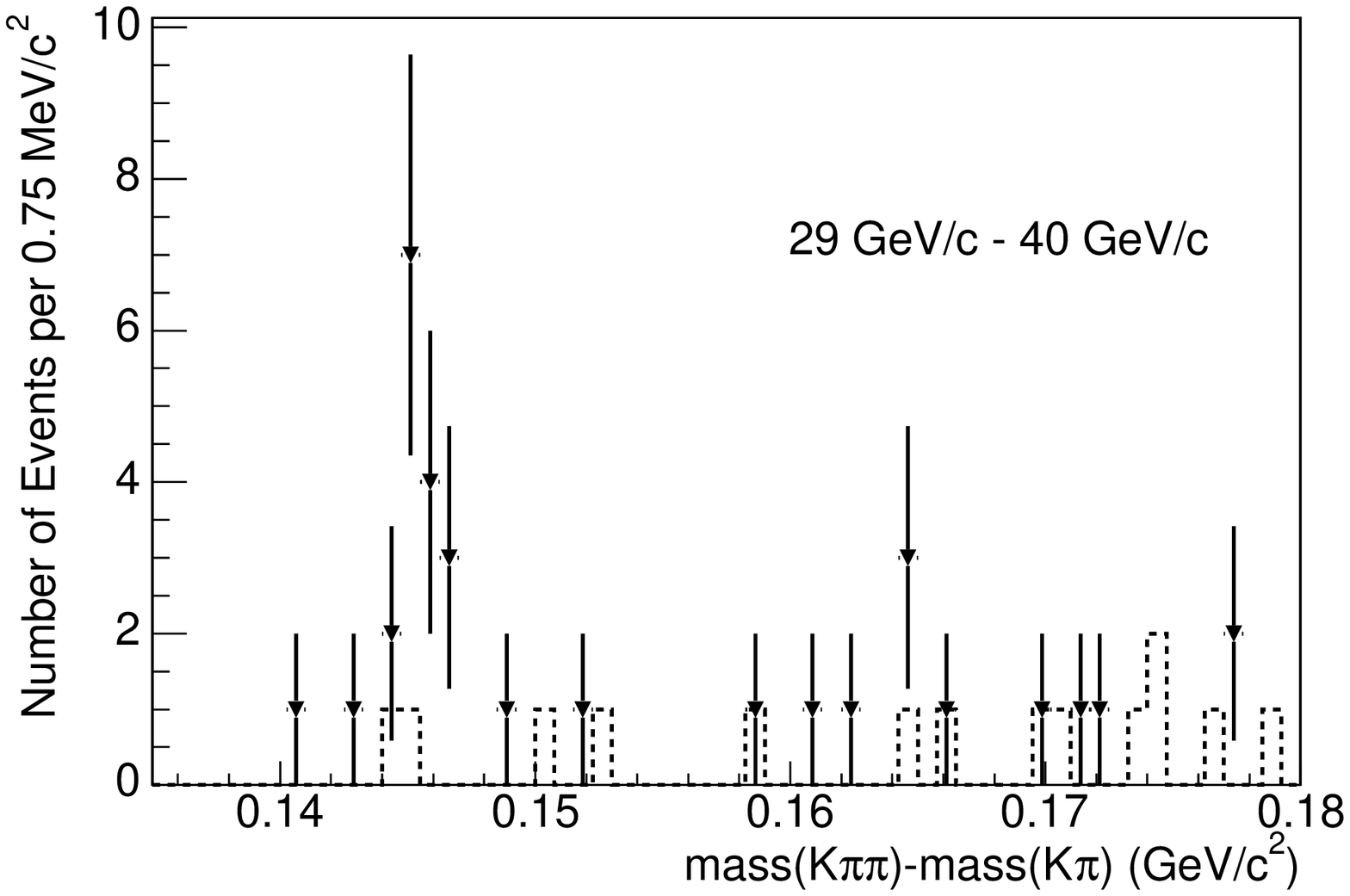,width=0.48\textwidth}
\psfig{figure=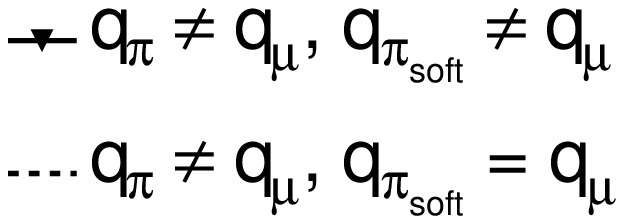,width=0.48\textwidth}
\caption{The mass difference between
$\pi_{soft} \pi^+ K^- $ and $\pi^+ K^-$  for 
different ranges of $p_T(\mu^- K^-\pi^+)$.
All events shown are selected such that 
the muon and the pion from the candidate  $D^0$ have 
opposite charge.  Points show events where the muon and soft pion have opposite charge, which is the right-sign correlation, 
and the dashed histograms represent 
events where the muon and soft pion have the
same charge, which is the wrong-sign correlation.  The $\chi^2$ per degree of freedom for these fits ranges from 1.1 to 1.3. Because of low statistics, 
the yield in the highest $p_T$ bin is measured by 
sideband subtraction.\label{Dstarfitmassplot}}
\end{figure}

\begin{figure}
\psfig{figure=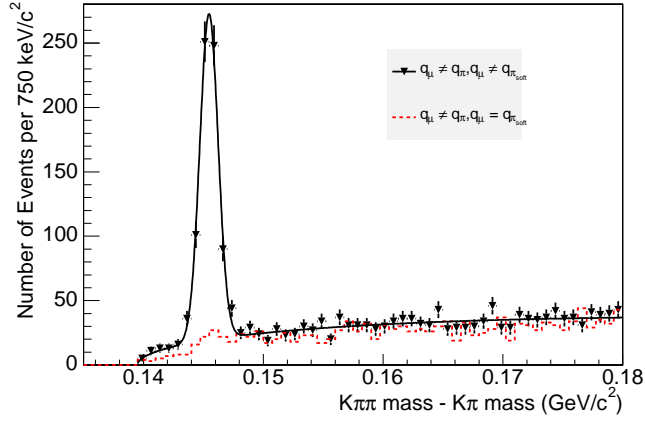,width=0.50\textwidth}
\caption{Distribution of the difference between the  $\pi_{soft}^+ \pi^+ K^-$ 
and the $\pi^+ K^-$ invariant masses for $p_T(\mu^- K^- \pi^+)>$ 9.0 GeV/$c$.  The solid histogram indicate events with the right-sign correlation between muon and pion, and the dashed histogram contain events where the muon and pion have the wrong-sign correlation.\label{DstarfitmassplotTotalNoSVT}}
\end{figure}

For the decay ${D^*}^+ \to D^{0} \pi^+$, 
$D^0 \to K^- \pi^+$, the muon and $D^0$ reconstruction is
the same as described above. 
We additionally require the presence of the pion from the
${D^*}^+\to D^0 \pi^+$ decay with $p_T> 400$ MeV/$c$ and
$|\eta|<1.0$ that passes COT fiducial and quality 
selection criteria.  
Since this pion tends to have 
low $p_T$, we refer to this as the ``soft pion''
($\pi_{soft}$.)  We require that the muon and soft pion 
have opposite charges and only consider events where 
the $K^- \pi^+$ invariant mass is between $1.82 \, {\rm GeV}/c^2$ 
and $1.90 \, {\rm  GeV}/c^2$.  As can be seen in 
Fig.~\ref{D0fitmassplot}, this mass window is wide 
enough so that no real $D^0$ decays are lost.  We require
$|\Delta z_0|$ between any two 
of the decay products of the $H_b$ must be less than 
$5\, \rm cm$.  In order to minimize the background, we look at the 
mass difference $\Delta m$ between the $K^- \pi^+ \pi_{soft}^+$ and the
$K^- \pi^+$.  We fit the mass difference distribution 
to a Gaussian plus the background shape 
$a \sqrt{\Delta m - m_{\pi}} e^{b (\Delta m - m_{\pi})}$, 
where $a$ and $b$ are free parameters in the fit to the data.
The mass difference plots 
are shown in Fig.~\ref{Dstarfitmassplot}.  In 
order to make comparisons between the two measurements, 
we present the data in $p_T$ bins of 
the $\mu^- D^0$ system.  The yields are included in 
Table~\ref{yieldtable}.  Figure~\ref{DstarfitmassplotTotalNoSVT} 
shows the mass difference 
for all events with $p_T(\mu^- D^0)>$ 9.0 GeV/$c$ 
and  $|\eta(\mu^- D^0)|< 0.6$. 

\section{Efficiency Measurements}
\label{sec:eff}

The efficiencies of the CDF II trigger and detector 
components for the cross-section measurement are determined
whenever possible by using data collected by complementary 
trigger paths.  The trigger paths utilized are described 
in Section~\ref{sec:trigger}.

We have separated the overall efficiency, 
$\epsilon$, into nine separate 
measurements, denoted $\epsilon_1$-$\epsilon_9$.  
The first eight terms are extracted using 
relative efficiencies and the ninth term is measured as 
an 
absolute efficiency.  
To calculate $\epsilon_8$, we
measure the efficiency for events to
pass selection criterion eight \it relative \rm to
events that already have passed criterion nine. 
This yields $\epsilon_8 = \epsilon^{rel}_8 \epsilon_9$.
Repeating this procedure for selection criterion seven, we
calculate the relative efficiency for events to pass
selection seven relative to events that have already passed
selections eight and nine: $\epsilon_7 = \epsilon^{rel}_7 
\epsilon_8$.   Repeating this process for each of
the selection criterion, we arrive at a final efficiency
which is

\begin{equation}
\epsilon = \epsilon_9 \prod_1^8 {\epsilon^{rel}_i}.
\end{equation}
The absolute efficiency measurement ($\epsilon_9$) is 
the efficiency for finding a charged-particle 
track in the COT.  This 
absolute efficiency is determined
using a combination of data and
MC simulation based upon \sc geant \rm \cite{geant}.  
A brief description of 
these efficiency measurements is provided below. Details
of all of the efficiency measurements and 
parameterizations may be found in Ref.~\cite{thesis}.
We perform these  measurements in bins of
kinematic variables.  For example, the level~1 trigger
efficiency depends upon the momentum of the muon.  The
efficiency corrections are then applied according to 
the kinematics of each event.  


\subsection{Level 1 Efficiency ($\epsilon_1$)}
For the $\mu$SVT trigger path, the level~1 muon trigger
requirement consists of an XFT track with $p_T>4\, {\rm GeV}/c$ that
is matched to track segments in the CMU and CMP subdetectors.
We 
measure the efficiency of this trigger with respect to the offline 
CMUP muon reconstruction efficiency.  To do this, 
we reconstruct the $J/\psi \to \mu^+ \mu^-$ signal in events that were 
collected via the inclusive muon trigger path.  The trigger
muon ($p_T > 8\, {\rm GeV}/c$) in these events
is biased for the efficiency measurement, but the other 
decay muon of the $J/\psi $ can be used to determine the 
trigger efficiency.  We refer to the $J/\psi$ decay muon
that is  unbiased by the trigger 
as the ``probe" muon, since it can be used
to directly measure the trigger efficiency.  We then compare the 
number of times the probe muon satisfied the trigger to the number of
probe muons that were within the acceptance of the trigger.
 
We require both muons of the $J/\psi $ 
to pass COT, CMU, and CMP fiducial and quality selection criteria.  
The probe muon must also have $|\eta| < 0.6$ and 
$p_T > $ 4 GeV/$c$.  We find a signal yield of
approximately 1900 events. 
Sideband subtraction is 
used to parameterize the efficiency 
in terms of muon $p_T$ and $\eta$.  We find the plateau
efficiency ($p_T\gtsim 5\, \rm GeV/{\it c}$) to be
$90\%$.

\subsection{Level 2 XFT Efficiency ($\epsilon_2$)}

For the $\mu$SVT signal path, the XFT tracks found in 
level~1 are used as input for the level~2 SVT trigger.
The trigger path requires at least one XFT track, not associated
with the muon candidate, having $p_T>2\, {\rm GeV}/c$; the SVT further 
requires that the impact parameter for this track 
be between $120 \mu \, \rm m$ and $1\, \rm mm$. The overall
efficiency of this selection is given by the product of
the XFT and SVT efficiency.  We require that either the kaon 
or pion
from the $D^0$ decay satisfy this selection.
 

The XFT efficiency is measured for 
pions and kaons with $p_T>$ 2.0 GeV/$c$ and  
$|\eta|<1.0$.  Given their different ionization energy losses,
we treat kaons and pions separately in this measurement.
In 
order to obtain pure samples of kaons and pions, we reconstruct 
$D^+ \to K^- \pi^+ \pi^+$ events, $D^+ \to \phi \pi^+$, 
$\phi \to K^+ K^-$ events, and $D_s^+ \to \phi \pi^+$, 
$\phi \to K^+ K^-$ events that were collected 
by the heavy flavor trigger path.
We have roughly 200 000, 
9500, and 18 000 signal events respectively in the
three modes.
Two of the tracks from each decay are required 
to pass the trigger requirements, and the 
third track is used to probe the XFT efficiency.  
We 
apply  both COT and SVX~II fiducial and quality selection criteria on the 
tracks that fired the trigger, and COT fiducial and 
quality selection criteria on the tracks used to probe the XFT efficiency.  
For the resonant decays, a cut is placed at $\pm 10\, \rm MeV$
around the 
$\phi$ mass.  Sideband subtraction is used to 
measure the  pion and kaon yields.  We parameterize the 
efficiencies in terms of $p_T$, $\eta$ and $\phi$.  We additionally parameterize
the XFT efficiencies over the time that the data was acquired. 
The average efficiencies are approximately
$90\%$ for pions and $85\%$ for kaons.

\subsection{Level 2 SVT Efficiency ($\epsilon_3$)}

The second component of the level 2 trigger 
efficiency is 
the SVT efficiency for 
tracks that have already been identified by the XFT.  The SVT 
efficiency does not depend on the particle species, so 
we measure the efficiency using a high statistics sample 
of $J/\psi \to \mu^+\mu^-$ 
events that were collected 
by the inclusive $p_T > 8\, {\rm GeV}/c$ muon trigger
path. The technique here is similar to the technique utilized
to measure the level~1 efficiency.
 
We use tracks that have 
$p_T>$ 2.0 GeV/$c$, $|\eta|<1.0$, 
and $120\mbox{ }\mu \mbox{m }<|d_0  
|< 850\mbox{ }\mu \mbox{m}$.  The muon from the $J/\psi$ that
did not satisfy the trigger is used as the probe track.  On events 
where both muons satisfy the $p_T>8\, {\rm GeV}/c$ inclusive muon trigger,
both muons are used as probe tracks. 
We impose COT 
and SVX~II fiducial and quality selection criteria on the probe track, 
and require the track to be 
matched to an XFT track.  Therefore the SVT efficiency
is measured relative to a track found offline with
the required number of SVX~II hits.  There are 
approximately 71 000 
tracks that pass these selection criteria.  The efficiency is 
parameterized in terms of $p_T$, $d_0$, and track isolation,
which is a measure of the number of tracks found within an 
angular region of $\delta\phi < 5^\circ$.
For tracks with $p_T \gtsim 3\, {\rm GeV}/c$, the plateau
efficiency for the SVT is approximately $90\%$.  The $\eta$ dependence 
of the efficiency is small, and further mitigated because the tracks 
used in the efficiency measurement sample the detector in a manner similar 
to the signal.  Since the SVX~II hit efficiency is not part of the 
level 2 SVT efficiency calculation
(it contributes to the SVX~II efficiency described below) the
inefficiency here is dominated by the track-finding and fitting algorithms
utilized in the SVT.

The track isolation dependence of the SVT efficiency requires
special treatment.  The occupancy in the silicon detector arises
from a number of sources, including very low momentum tracks from
the underlying $p\overline{p}$ interaction along with a 
spectrum of 
tracks originating from multiple $p\overline{p}$ interactions.
These effects are not modeled reliably 
in the Monte Carlo simulation, and so we reweight the data
based on
the isolation of the signal events.  In addition to the 
raw yields, Table~\ref{yieldtable} also 
lists the yields after correction for the SVT efficiency.

\subsection{Level~3 Efficiency ($\epsilon_4$)}
At level~3, our trigger path requires that a 
muon with $p_T > 4 {\rm GeV}/c$ be reconstructed.  In addition, a track
identified in the COT (other than the muon)
must be matched
to an SVT track meeting the level~2 trigger 
requirements.  
The muon track and COT-SVT track must
have an invariant mass of less 
than $5.0 \, {\rm GeV}/c^2$ and $2.0^\circ<\Delta \phi_0 < 90^\circ$.  
To find the efficiency of this trigger, we 
look at $\sim\! 30\  000$ events that 
were acquired on the $H_b$ semileptonic backup path.    
We parameterize the efficiency in 
terms of the $p_T$ of the muon and the $p_T$ of the SVT track.
For muons with $p_T\gtsim 4.5\, {\rm GeV}/c$, the level~3 
efficiency is approximately $97\%$.

\subsection{SVX~II Efficiency for Triggered Pions and Kaons ($\epsilon_5$)}

In order to determine the efficiency for attaching 
SVX~II hits to a COT track, we look at muon tracks from 
$J/\psi \to \mu^+ \mu^-$ decays collected 
on the $J/\psi$ dimuon trigger path.  We use 
this sample in order to have a large sample of tracks 
with negligible fake rate.  
The tracks are required to pass COT quality and 
fiducial, as well as SVX~II, selection criteria.  We consider 
only tracks 
that are expected to pass through the four layers 
of the SVX~II used in the SVT trigger.  The average efficiency 
is observed to be $80\%$, where the inefficiency is dominated 
by dead silicon ladders and poor efficiency near the ends of
the silicon ladders. 
To account for the 
dependence on SVX~II coverage, we correct the 
efficiency as a function of the $z$ 
position of the track at its origin.

\subsection{SVX~II Efficiency for Untriggered Pions and Kaons ($\epsilon_6$)}

We require hits in at least three layers of the SVX~II 
on the $D^0$ decay track that is not required by the trigger.
We find the efficiency by looking in our signal sample with and without 
the SVX~II hit requirement 
for the second track.  This yields an SVX~II efficiency for
the untriggered track of 
$(93.2 \pm 1.0)\% $.
The efficiency for the untriggered track
is higher than that for the triggered track because
of the looser requirement on the number of SVX~II hits.

\subsection{CMU efficiency ($\epsilon_7$)}
In order to find the efficiency for linking a 
muon COT track to a CMU track segment, 
we reconstruct $J/\psi \to \mu^+ \mu^-$
events using both the inclusive muon and 
semileptonic backup trigger paths. 
The muon tracks  
must pass COT fiducial and quality selection criteria.  One of the decay muons
from the $J/\psi$ must satisfy the muon 
trigger requirements.  For events acquired on the semileptonic
backup trigger path, the probe 
track must satisfy the SVT trigger requirements.  The probe tracks are extrapolated 
to the CMU, and are required to pass CMU fiducial selection 
criteria.  The probe tracks must have 
$p_T > 4.0\, {\rm GeV}/c$ and 
$|\eta|<0.6$.    
The efficiency is found to be 
independent of $p_T$, $\phi$, and $\eta$, and 
so we use a single efficiency 
value of  $(79.5 \pm 1.3)\%$.

\subsection{CMP efficiency ($\epsilon_8$)}
We find the efficiency for linking a CMP track segment 
to a COT track that has already been linked to a CMU track segment by 
reconstructing $J/\psi \to \mu^+ \mu^-$ candidates in events collected by the $J/\psi$ dimuon trigger path.  This trigger 
path does not require hits in the CMP, so either decay muon from 
the $J/\psi$ can be used to probe the CMP efficiency.  
We require the muon to pass COT and CMU fiducial and quality 
selection criteria.  The muon is extrapolated out to the CMP 
and required to pass CMP fiducial selection criteria.  
We examine only muons with $p_T>$ 4.0 GeV/$c$ and $|\eta|<0.6$.  
The efficiency is parameterized in terms of 
$p_T$ and $\eta$.  The average CMP efficiency is $85\%$.

\subsection{COT efficiency ($\epsilon_9$) \label{COTeffSection} }

To determine the absolute COT track-finding efficiency, we
use a combination of data along with a 
detailed MC simulation of the CDF~II tracking system.  The techniques
described here have been utilized in previous CDF analyses 
\cite{AcostaJpsi2005, chun}.   We have verified that the MC 
simulation accurate models the charged-particle kinematics of our signal
sample \cite{thesis}.  We have additionally performed detailed studies
of the COT simulation and find that the data is well-modeled by the
simulation \cite{chun-thesis}.

First, we use a MC simulation only technique to 
compare charged particles generated by the \sc pythia \rm 
MC program to charged tracks reconstructed after full MC
detector simulation and event reconstruction.  This yields
our measurement of absolute COT track-finding efficiency.  

To assess the systematic uncertainty on this measurement, we
use a combination of data and MC simulation to ``embed'' simulated
MC kaons, pions and muons into data events acquired via the inclusive
muon trigger path.   The MC simulated COT hits from the simulated
track are incorporated into the detector data from a real event.  
To accurately emulate track reconstruction in $\mu^- D^0$ events,
the embedded MC simulated track is placed within a 
region  in $\eta$-$\phi$ space that is
within $\sqrt{\delta \eta^2 + \delta \phi^2}<1.4$ of the 
trigger muon.  
The hybrid event, which consists of 
data plus hits from a single simulated track,
is
then reconstructed in the same way as our signal sample.  The 
simulated track is considered to be correctly found if a 
reconstructed track is matched closely in $p_T$ and $\phi$ 
to the simulated track
embedded in the event.

We vary the matching criteria to assess the rate at which 
tracks may be found but reconstructed with incorrect 
parameters.  For tracks with 
$p_T > 1.0\, {\rm GeV}/c$ (which includes all muons and kaons, and 
the pions from the decay of the $D^0$), the 
track-finding 
efficiencies derived from MC-only simulation and  track embedding
agree to within $1\%$ and are insensitive to the details of the 
matching criteria.
The COT efficiency
for muons is approximately $99\%$.  For pions and kaons with
$p_T > 2\, {\rm GeV}/c$ the COT efficiency is greater than $90\%$.
For pions with $p_T < 1.0\, {\rm GeV}/c$, which are used in reconstructing 
the ${D^*}^+ \to D^{0} \pi^+$ decay, we observe a significant difference
in efficiency estimates from the different techniques, and we use these
variations to assess a systematic uncertainty on the low $p_T$ 
efficiency measurement.
It is additionally necessary to account for uncertainty in the
amount of material in the inner detector region, which also affects
the absolute track-finding efficiency.  We make a 
conservative estimate of the
systematic uncertainty 
by varying the total amount of material by 25\%.  This
rather large variation in the amount of passive material in the detector
translates into a systematic uncertainty on the tracking efficiency of 
approximately $2.6\%$.

\subsection{Efficiency Summary}

For a given signal event that is within the acceptance of 
the detector, the absolute efficiency to reconstruct the event is
the product of the efficiency terms outlined in this section.  In 
many cases, the efficiency depends upon the track momentum as well as its
position within the detector.  The details of these efficiency
parameterizations can be found in Ref.~\cite{thesis}.   For signal
events that are well away from turn-on effects, the overall 
trigger and reconstruction efficiency is approximately $35\% $.

\section{Acceptance}
\label{sec:acc}

Our detector acceptance is defined by the selection criteria and the
active regions of the detector that cover those criteria.  
We determine the detector acceptance from MC simulation by first 
generating
single $b$ hadrons according to measured momentum and rapidity 
spectra.  These hadrons are then decayed using the \sc evtgen \rm 
package \cite{evtgen} and then fed into a \sc geant \rm simulation of the
detector.  The MC simulation of the detector is utilized to map out the
active regions of the detector.  At this stage, we are not using the 
simulation to quantify detector inefficiencies, since those are measured 
using techniques described in the previous section. 
We generate $b$ hadrons based upon two distinct distributions in
transverse momentum and rapidity.   
The first is based on NLO \cite{NDE1,NDE2} QCD theory 
with MRSD$_0$ parton distribution
function \cite{mrsd0}, a $b$ quark mass of $4.75\, {\rm  GeV}/c^2$, and renormalization and factorization scales equal to 
$\sqrt{m_b^2+p_T^2}$.  Peterson fragmentation is used, with $\epsilon_P = 0.006$.  The other spectrum is based 
on the measured $H_b \to J/\psi X$ cross-section \cite{AcostaJpsi2005}, which
we refer to below as the ``CDF MC sample.''  We use the production fractions 
$f_u : f_d : f_s $ of 1:1:0.270 \cite{PDG}.  We assume that 
$\Lambda_b$ decays provide a negligible contribution to our
signal because of the low probability for a semleptonic $b$-baryon
decay that would also yield a $D^0$ meson.

\begin{figure}
\psfig{figure=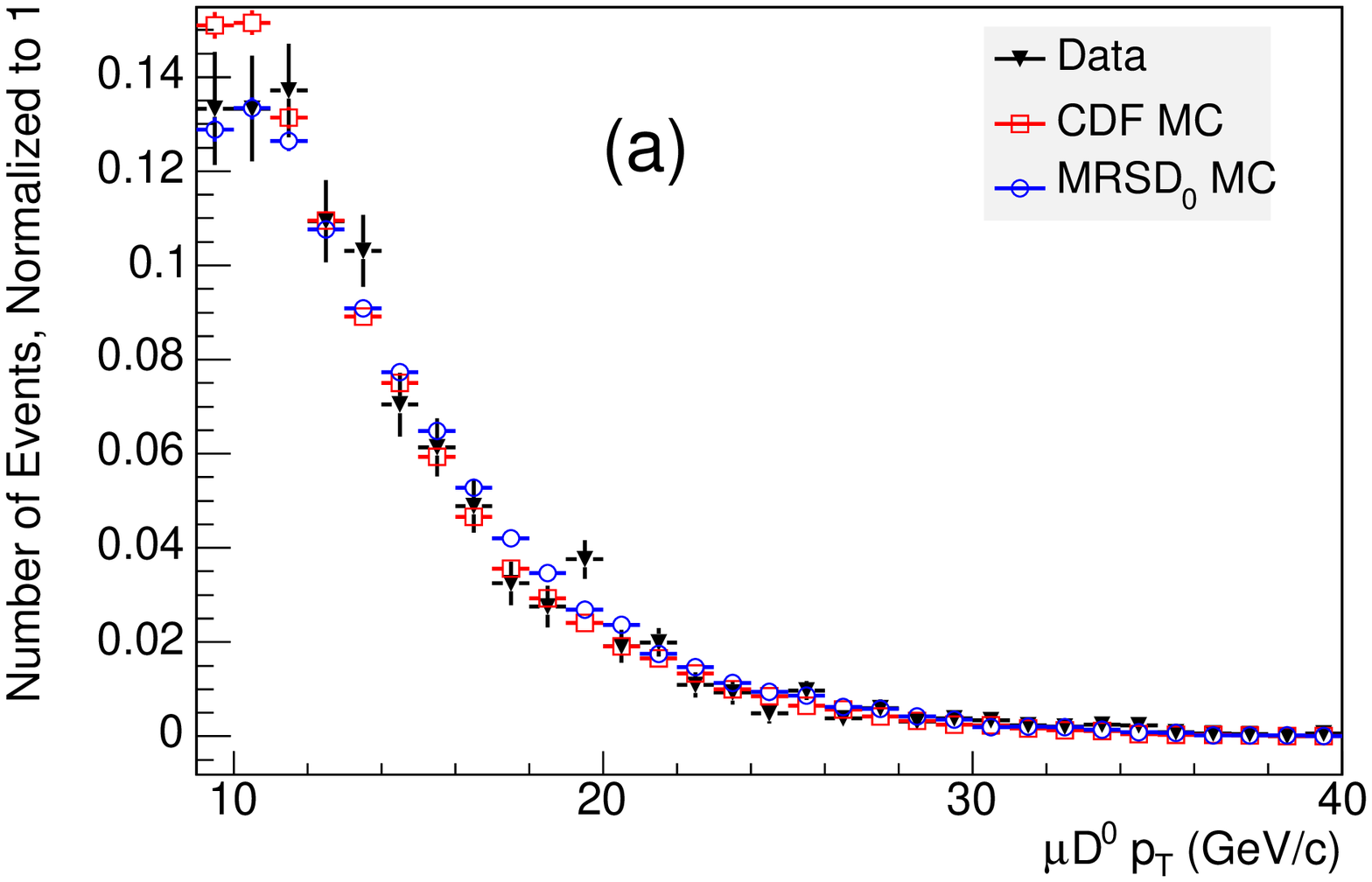,width=0.45\textwidth}
\psfig{figure=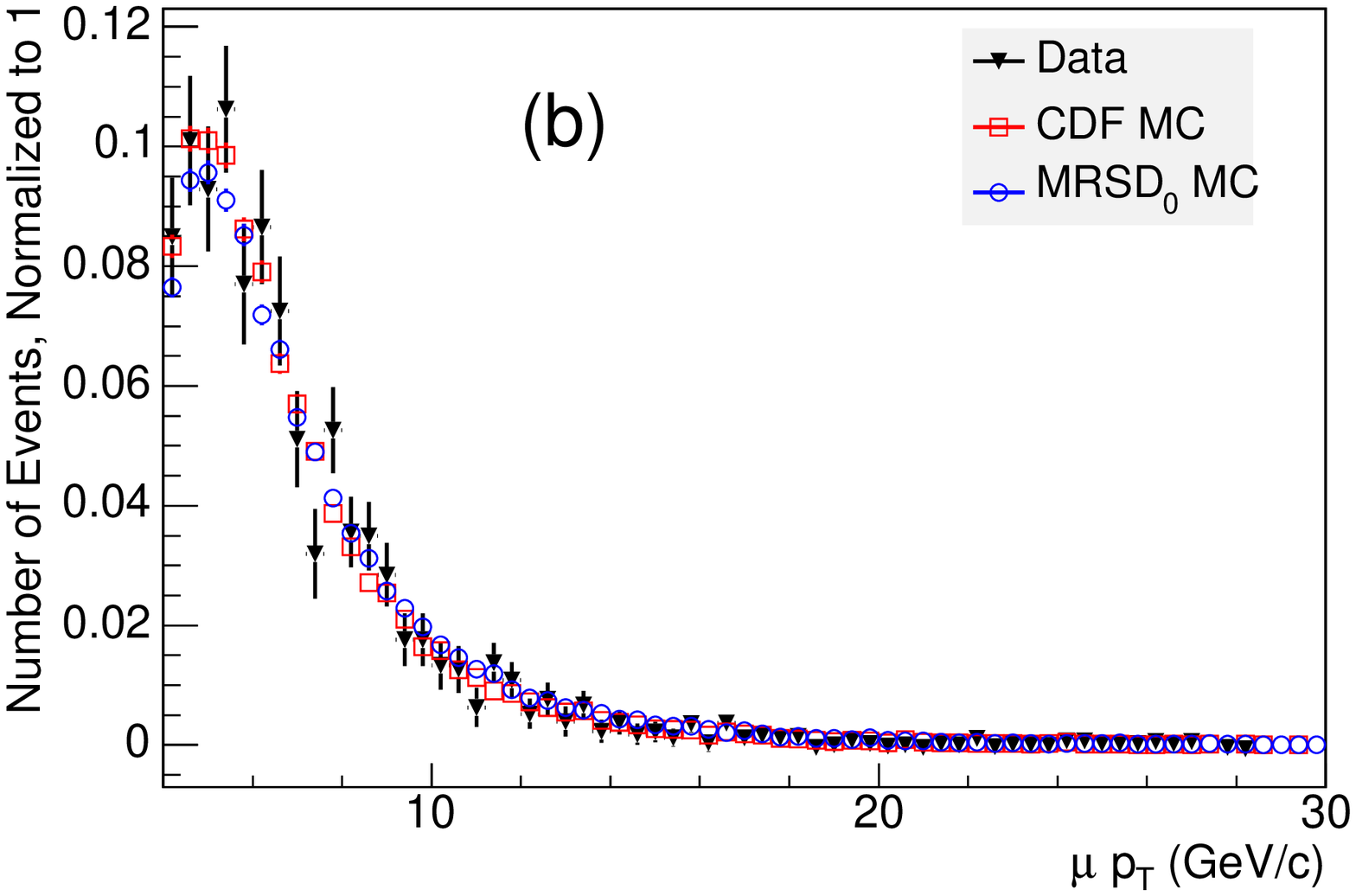,width=0.45\textwidth}
\psfig{figure=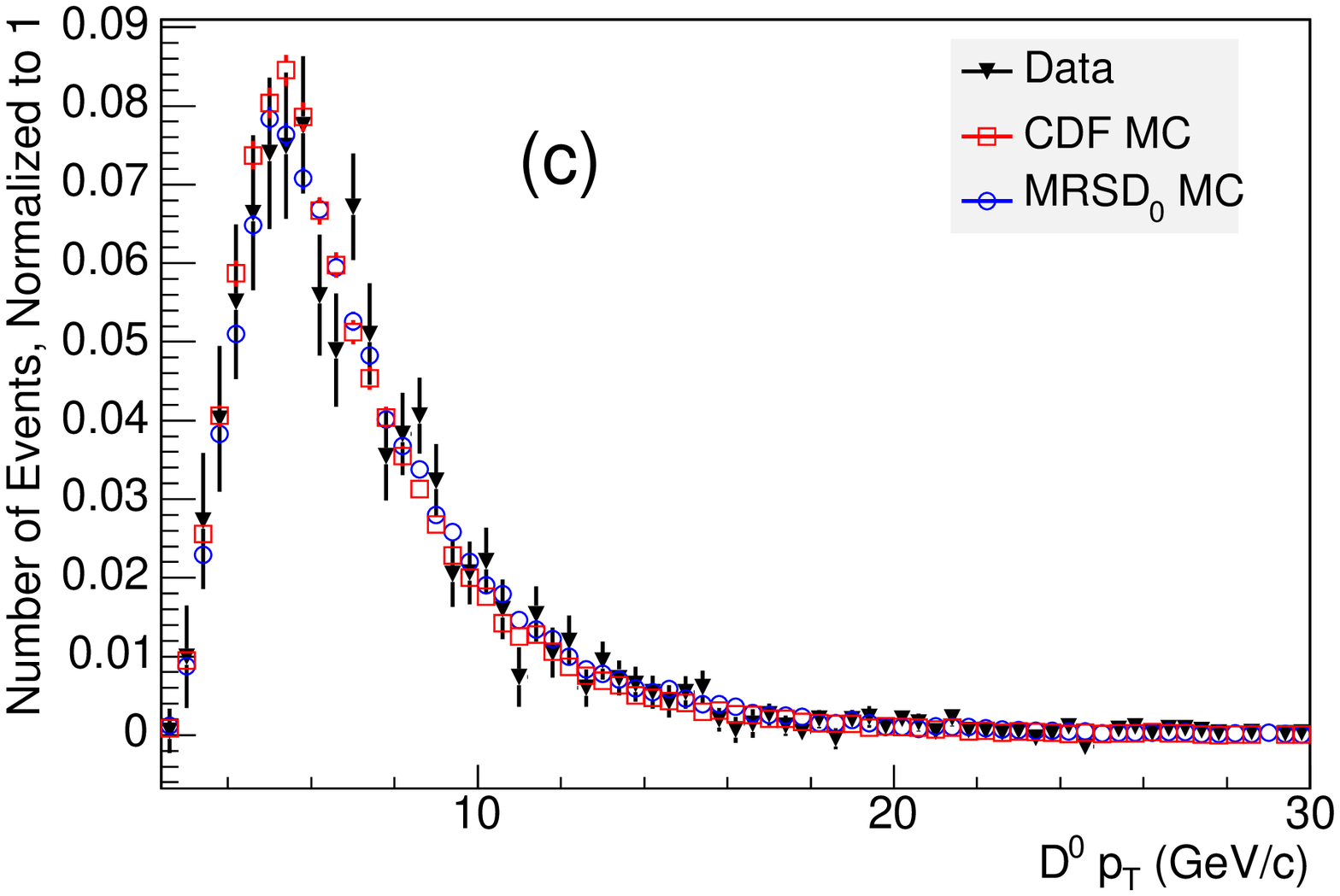,width=0.45\textwidth}
\caption{A comparison of the two MC simulated 
samples to data, plotted as a function of 
the $p_T$ of (a) the $\mu^- D^0$, (b) the $\mu$, and (c) the $D^0$.  The 
data is seen to have a higher $p_T(\mu^- D^0)$ than 
the CDF MC sample, while the data is seen
to have a lower $p_T(\mu^- D^0)$ than the 
MRSD$_0$ NLO MC simulation. \label{DatatoMCcomppt}}
\end{figure}

To demonstrate the degree to which the MC simulation reproduces the data, 
Figs.~\ref{DatatoMCcomppt}(a,b,c) show the $p_T$ distributions of the 
$\mu^- D^0$, $\mu$, and $D^0$ for the MC simulation 
after all selection criteria and efficiencies 
have been applied, and 
for data after all selection criteria and corrections for the SVT efficiency.  
The data are seen to have higher $p_T(\mu^- D^0)$  
than the MC simulation based upon the $H_b \to J/\psi X$ cross-section 
result \cite{AcostaJpsi2005}, 
while they are seen to have
lower $p_T(\mu^- D^0)$  than the 
spectrum from the MRSD$_0$ MC simulation, referred to below
as the ``MRSD$_0$ MC sample.''  We 
therefore measure the cross sections by 
taking the average of the cross sections 
from the two methods  and assign one-half the 
difference as a systematic uncertainty.  

We perform additional comparisons between the data and the 
two MC simulated samples.  Distributions 
such as $\eta(\mu^-D^0)$, $p_T(\mu^-)$, and $p_T(D^0)$ are all
seen to agree well between the data and both MC simulated
samples.  These
comparisons can be found in Ref.~\cite{thesis}.
Table~\ref{acctimeseff} shows the 
acceptance times efficiency (excluding the SVT efficiency) 
as a function of  $p_T(H_b)$ bin calculated 
using the CDF and MRSD$_0$ samples \cite{thesis}.

There is significant uncertainty in the branching ratios for
$H_b\to \mu^- \rightarrow D^{**}X$ where the $D^{**}$ decays to
$D^0 X$.  An incorrect modelling of this fraction can affect the
$p_T(\mu^- D^0)$ spectrum and therefore the acceptance.  To 
conservatively assess the systematic uncertainty on the measured
cross section, we modify $\mathcal{B}(H_b \to
\mu^- D^{**})$ by $\pm 50\%$.

\section{Backgrounds}
\label{sec:bkg}

While fitting the $D^0$ peak to a Gaussian allows us to determine how many 
$\mu^- D^{0}$ 
events are in our sample and account for noncharm background, it is possible that there 
are real $\mu^- D^0$ events that are not part of our $H_b \to \mu^- D^0$ signal.  One source 
of this background is direct $c\overline{c}$ production.  It is possible for direct 
charm to mimic our signal by having one charm quark decay to a muon, 
and the other charm 
quark fragment into a $D^0$.  The second background source is $b\overline{b}$ production, 
where one $b$ hadron decays to $D^0 X$, and the other $b$ hadron 
follows the decay chain 
$H_b \to D \to \mu^- X$.  Since these background sources are irreducible,  
we estimate their contributions and correct the total number of
observed signal events \cite{thesis}.

\subsection{$c\overline{c}$ Background Estimate}

\begin{figure}
\begin{center}
\psfig{figure=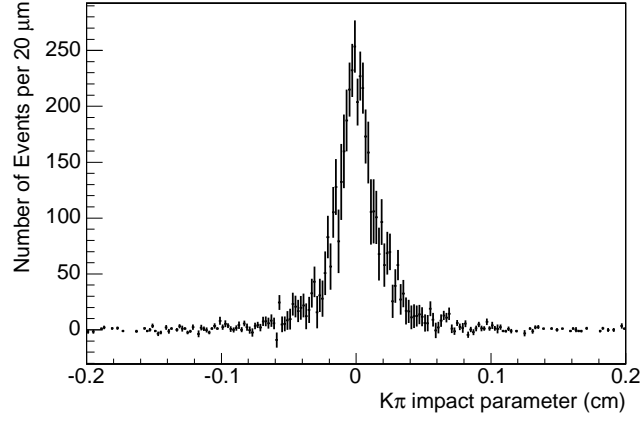,width=0.50\textwidth}
\end{center}
\caption{The $D^0$ impact parameter distribution for the data after sideband subtraction. \label{D0D0dat}}
\end{figure}

In order to estimate the fraction of events in our sample from direct charm, we use the 
impact parameter of the $D^0$ \cite{chun}.  As the $D^0$ mesons from direct charm are created at the 
primary vertex, they should point back to the primary vertex
within an error determined by the detector 
resolution.  The $D^0$ from $b$ decays, on the other hand, are created at the secondary 
$b$ vertex, and are less likely to point back at the primary vertex.  
The $D^0$ impact parameter distribution for our signal sample is shown
in Fig.~\ref{D0D0dat}.  Although dominated by $D^0$ from $b$-hadron
decays, this distribution contains a prompt component arising from 
direct $c\overline{c}$ production.  We use MC simulation to determine 
the expected impact parameter distributions for $D^0$ mesons from $b$-hadron
production and direct production.

To determine the impact 
parameter distribution for $D^0$ mesons from $b$ decays, we use the same MC 
simulated sample that 
is used to find the acceptance above.  In order to get the $d_0$ distribution of the 
$D^{0}$ from direct charm, we generate $D^0 \to K^- \pi^+$ events originating from the 
primary vertex.  The generated prompt $D^0\to K^- \pi^+$ sample  
has lower $p_T(D^0)$ than does
our signal sample, so we reweight the 
$p_T$ spectrum of the MC simulated 
$D^0$ so that it matches the $p_T(D^0)$ spectrum 
in $H_b\rightarrow \mu^- D^0$ events from the \sc pythia \rm  MC 
simulation \cite{pythia}.

In the direct charm and $H_b\rightarrow D^0$ samples, we calculate the impact parameter using 
generator level MC simulated quantities.  In order to estimate the uncertainty on the $D^0$ impact parameter due 
to detector resolution, we look at the sideband subtracted $D^0$ impact parameter uncertainty 
distribution in our signal sample acquired on the $\mu$SVT trigger.   We see that the average uncertainty is 
approximately 34 $\mu \rm m$, so we smear the MC simulated 
impact parameters using a Gaussian 
resolution function with a standard deviation of 34~$\mu \, \rm m$.  We do a bin-by-bin $\chi^2$ fit to determine the ratio of events from 
direct charm to those from $b$ hadron decays.  We find that using the $d_0$ distribution 
from the $H_b$ events generated with the $p_T$ spectrum measured in the
CDF $H_b\rightarrow J/\psi X$ cross section 
measurement \cite{AcostaJpsi2005}, we measure 
a charm fraction of 
$(6.3 \pm 2.1)\%$, while using 
the $d_0$ distribution from the $b$ events in the 
MRSD$_0$ sample gives a charm fraction of $(5.4 \pm 2.1)\%$.  
So as a final result, we use
the average charm 
fraction of $(5.9 \pm 2.1{\rm (stat)} \pm 0.4 {\rm (syst)})\%$.
 
For the differential cross section, we must also account for the
$p_T$ spectrum of $\mu^- D^0$ candidates arising from direct
$c\overline{c}$ production.  For normalization, we use the
total $c\overline{c}$ yield as calculated above.  To determine
the $p_T(D^0)$ spectrum, we use MC simulation, where the generated
events are reweighted to match the predicted direct charm spectrum
\cite{cacciari}.  From this reweighted MC sample, we estimate the
bin-by-bin contribution from direct $c\overline{c}$ background. 

\subsection{$b\overline{b}$ Background Estimate}
The presence of $b\overline{b}$ events can mimic the signal in
the cases where the $\mu^-$ and $D^0$ come from different 
$b$ hadrons.  An example of this is where the $b$ hadron decays as
$b\to D^0X$ and the other follows the decay chain 
$\bar{b}\to \bar{c}\to \mu^-X$.
This configuration
provides a 
$\mu^-D^0$ candidate 
with the proper charge correlation, constituting an 
irreducible background to genuine $H_b \to \mu^{-} D^0$ decays.


To investigate this background source, we look at 
events where one $b$ hadron decays to a $D^0$ meson,
and the other follows the decay chain 
$\overline{b}\rightarrow \mu^+ X$,
providing the wrong-sign $\mu^+ D^0$ combination.  
The right-sign $b\overline{b}$
background is then estimated by 
counting wrong-sign $\mu^+ D^0$
events and correcting this by the expected ratio
of the two decay chains  
$\overline{b}\rightarrow \mu^+$ and $\overline{b}\rightarrow
\overline{c}\rightarrow \mu^-$.

We begin by looking for wrong-sign $\mu^+ D^0$ events, where we have 
a $D^0 \to K^-\pi^{+} $ peak in events where the pion and muon have the same 
charge, $q_{\mu} = q_{\pi}$.  
This is complicated 
by two factors.  First, there 
is a 
large background in the wrong-sign sample from real
$\mu^-D^0$ signal events where the $D^0\rightarrow K^- \pi^+$
decay is incorrectly
reconstructed as  $\overline{D^0} 
\rightarrow K^+ \pi^- $.   
The reflection does not yield a narrow $D^0$ peak in the
pion-kaon invariant mass distribution
but instead yields a broad peak which we
parameterize using MC simulated $D^0$ decays.  
We use this parameterization and fix 
normalization of the reflection peak relative to
the signal peak based upon the MC simulation.
The MC simulated 
wrong-sign mass distribution is found to be 
insensitive to the input $p_T(b)$ spectrum.


The second complication
is that the doubly Cabibbo-suppressed decay of the 
$D^0 \to K^+\pi^{-} $ is also expected to yield a small peak in the
$K$-$\pi$ invariant mass distribution.  Based upon 
measured branching ratios \cite{PDG}, the number of 
doubly Cabibbo-suppressed events is expected to be
$0.0036 \pm 0.0003$ times the number of events in our signal peak.   
After correcting for these effects, 
we find a wrong-sign peak that is $(4.4 \pm 1.7)\%$ the size of the 
right-sign peak.

To convert this number into an estimate of the number of 
right-sign $b \overline{b}$ events,
we look at a generator level sample of $b \overline{b}$ MC simulation 
generated using \sc pythia \rm  
and decayed with \sc evtgen \rm using our best knowledge of branching ratios.  In this sample, we look for events where 
one $b$ decayed to a muon, and the other one produced a $D^0$.  In order to get a
sufficiently large
event sample, we use somewhat looser selection criteria on the $b \overline{b}$ than are in our sample, and 
do not apply any of our efficiency measurements to the events.  We find
the ratio to 
be $N(\bar{b}\to\bar{c}\to\mu^-)/N(\bar{b}\to\mu^-)=0.23 \pm 0.02\, {\rm (stat)}$.
To take into account the effect of the looser event selection criteria,
 we apply a 50\% 
systematic, giving a final ratio of $0.23 \pm 0.02 {\rm (stat)} 
\pm 0.11 {\rm (syst)}$.  
Combining this result with the wrong-sign contribution of  
$(4.4 \pm 1.7)\%$ yields a $b \overline{b}$ background 
fraction of $(1.0 \pm 0.40 \pm 0.48)\%$.

\subsection{$H_b \to DD$ and $H_b \to D \tau$ Backgrounds \label{btoddsection}}
In addition to backgrounds from $c \overline{c}$ and $b \overline{b}$ events, a background  
also comes from the decays of a single $b$ hadron.  
These events can occur when the decay 
$b \to c \overline{c} s$ or $b \to c \overline{c} d$ is followed by 
one of 
the charm quarks decaying to a muon, and the other decaying 
through $D^0 \to K^- \pi^+$.  These events 
also can come from a decay of $b \to c \tau^{-} \overline{\nu}_{\tau}$, where the tau lepton decays 
to a muon, and the $c$ decays through $D^0 \to K^- \pi^+$.
Because these events come from real $H_b$ decays, the impact parameter of the $D^0$ will not 
necessarily be consistent with production at the primary vertex. 
The muon and 
the pion from the decay will have opposite charges, as is the case in $b \to c \mu^{-} \overline{\nu}_{\mu}$
events.  We account for this contribution by retaining
events  with $H_b \to DD$ and $H_b \to D \tau$ decays
in the MC simulation utilized for the acceptance measurement and
correct for it in the unfolding procedure 
described in Sect.~\ref{sec:results}.
Since the branching fractions for some of these
decays are not well known, we allow the number of events from $H_b \to DD$
and
$H_b \to D \tau$ to vary by $\pm 50\%$, which translates in a $2.5\%$ 
systematic 
uncertainty on the measured cross section.


\subsection{Fake Muons}

In fitting the $D^0$ mass distribution, we account for
combinatoric background 
using the sidebands of the mass distribution.  This
technique is not available for hadrons that are 
misidentified as muons, so we must estimate
the contribution of fake muons independently.  

For $c\overline{c}$ background events described above, the
background is assessed by looking at the impact parameter
of the $D^0$, independent of whether the muon is real 
or fake.   Therefore, the only source of fake muons that
we must account for are fake muons reconstructed in 
association with real $H_b \rightarrow D^0 X$ decays.  To
evaluate this, we use a sample of $D^+\rightarrow K^- \pi^+\pi^+$
decays from the heavy flavor trigger path that is enhanced
in $H_b \rightarrow D^+ X$ decays.   We  measure the
rate of tracks near the $D^+$ which are fiducial to the
CMU and CMP detectors and have $p_T>3\, \rm GeV/{\it c}$.  
Because of their smaller nuclear interaction cross section, kaons 
are misidentified as muons at a higher rate than pions.  To 
conservatively assess a muon misidentification rate,  we
assume that all of these tracks are kaons.  We then apply a
$0.5\%$ 
probability that a fiducial kaon is reconstructed as a
muon, as measured in Ref.~\cite{bbbar}.  From
this, we determine that the rate of fake muons in our signal
sample is approximately $0.05\%$, small enough to neglect.

\section{$H_b \to \mu^- D^{0}X$ Cross Section \label{sec:results}}

In order to extract the $b$-hadron differential cross section from the 
measured $p_T(\mu^-D^0)$ distribution, we unfold the observed 
$p_T(H_b)$ distribution  
on the basis of the MC simulated signal.
We determine a weight $w_{ij}$ that is defined as

\begin{equation}
w_{ij} = \frac{N({[H_b]}_i \to {[\mu^-D^0]}_j)}
{N({[\mu^- D^0]}_j)},
\end{equation}
where $N({[H_b]}_i \to {[\mu^-D^0]}_j)$ is the number of $b$ hadrons with
$|y|<0.6$ in $p_T(H_b)$ bin $i$ decaying to $\mu^- D^0$ in $p_T(\mu^-D^0)$ 
bin $j$, which pass all selection criteria, and $N({[\mu^- D^0]}_j)$ is
the number of $\mu^- D^0$ events in $p_T(\mu^- D^0)$ bin $j$, which pass
all selection criteria. 

We generate MC simulated $H_b$ events with $|y(H_b)|<0.8$ as described in
Section~\ref{sec:acc} to properly model
the detector acceptance.    
We also include events where the $\mu^- D^0$ is produced via a $H_b \to DD$
or $H_b \to D \tau$ decay in the denominator in order to take that background into account.  
For the determination of the weights $w_{ij}$, it is the 
shape of the $p_T(H_b)$ distribution, 
not its absolute normalization, that matters.  We list the weights 
from the CDF and MRSD$_0$ MC simulated 
samples in Tables~\ref{CDFweights} and~\ref{MRSD0weights}.  Then, to 
unfold the number of $b$ hadrons $N^{H_b}_{i}$ in a given
$p_T(H_b)$ bin $i$ we use the formula

\begin{equation}
N^{H_b}_{i} = \sum^N_{j = 1} w_{ij} N_{j}^{\mu^- D^0},
\end{equation}
where $N_{j}^{\mu^- D^0}$ is the number of 
events in $p_T(\mu^-D^0)$  bin $j$ from data, shown in Table~\ref{yieldtable}.
We now have all of the terms from Eq.~(\ref{CrossSectionEquation}) required to extract the
cross section. 


The statistical uncertainty on the cross section for 
$p_T(H_b)$ bin $i$ is given by
\begin{equation}
\delta^2_{stat}(\sigma_{i}) = (\delta N^{H_b}_i)^2 \left[\sum^N_{j = 1} 
\frac{w_{ij}}{\alpha_i\epsilon_i {\cal L}} \right]^2,
\end{equation}
where $\alpha_i$ and $\epsilon_i$ are the acceptance and efficiency for the 
bin $i$ respectively.  The statistical uncertainty on the total cross 
section is obtained by summing the uncertainties in quadrature over all
bins of $p_T(H_b)$.

We check the analytical calculation of the
statistical uncertainty using MC simulated
pseudo experiments which model
the signal and background distributions along with bin
migration of the signal.  We verify that the
MC simulation yields the same statistical uncertainty on the 
differential and integrated
cross section as we observe using the above analytical 
formula.  The uncertainty is calculated
by generating 1000 distinct MC simulated experiments corresponding to
the total sample $\mu^- D^0$ yield we observe.  The distribution of
fitted yields in each $p_T$ bin is observed to be Gaussian, and we
take the standard deviation of the distribution to be the statistical
uncertainty in each bin.

The systematic uncertainties are determined in general by 
varying the efficiency or acceptance 
in question by $\pm 1 \sigma$ and calculating the shift in the
measured cross section.   In the case of the 
efficiency measurements, which   
are
parameterized based upon fits to the data (\it e.g. \rm ,
the level 1 trigger 
efficiency is parameterized as a function of muon $p_T$ and $\eta$)
the variation in efficiency is taken from the
 error matrix determined in the 
fits.

Using this technique, the uncertainty derived based upon the
SVT efficiency fit appears to be underestimated.  The 
variation in the measured points is not reflected by the statistical
uncertainty in the fit for the SVT efficiency.  In this case, we
assess the systematic uncertainty by replacing the fitted efficiency
with the bin-by-bin points that are derived in the efficiency 
measurement.  The difference between the result and the 
value obtained from the SVT efficiency data 
points is taken as a fit modelling uncertainty and combined in quadrature with the 
1$\sigma$ systematic uncertainty described above 
to obtain the total systematic uncertainty \cite{thesis}.

The systematic uncertainty on the MC simulated $p_T$ spectrum is 
estimated by taking the fractional difference between the cross section found using the $H_b \rightarrow J/\psi X$
and the reweighted MRSD$_0$ NLO MC simulated samples.  The degree to 
which final state radiation affects the $D^0$ lineshape is assessed with
the \sc photos \rm MC simulation \cite{photos}.
Systematic uncertainty contributions are
shown in  
Table~\ref{SystematicsTable}.  It is worth noting that the systematic uncertainty arising from the 
MC simulated $p_T$ shape on the total
cross section is less than the systematic uncertainty due to the MC simulation
$p_T$ shape in most of the 
$p_T$ bins.  This is because uncertainty on the $p_T$ shape primarily affects 
the fraction of 
the total cross section in a specific $p_T$ bin.  When integrating over all bins, this effect is 
diluted.  Also, when finding the acceptance for a specific $p_T$ bin, one can only use the MC simulated events 
within that bin, while the overall measurement uses all of the MC simulated
events.  This leads to a lower 
statistical uncertainty in the MC simulated samples.

\begin{figure}
\psfig{figure=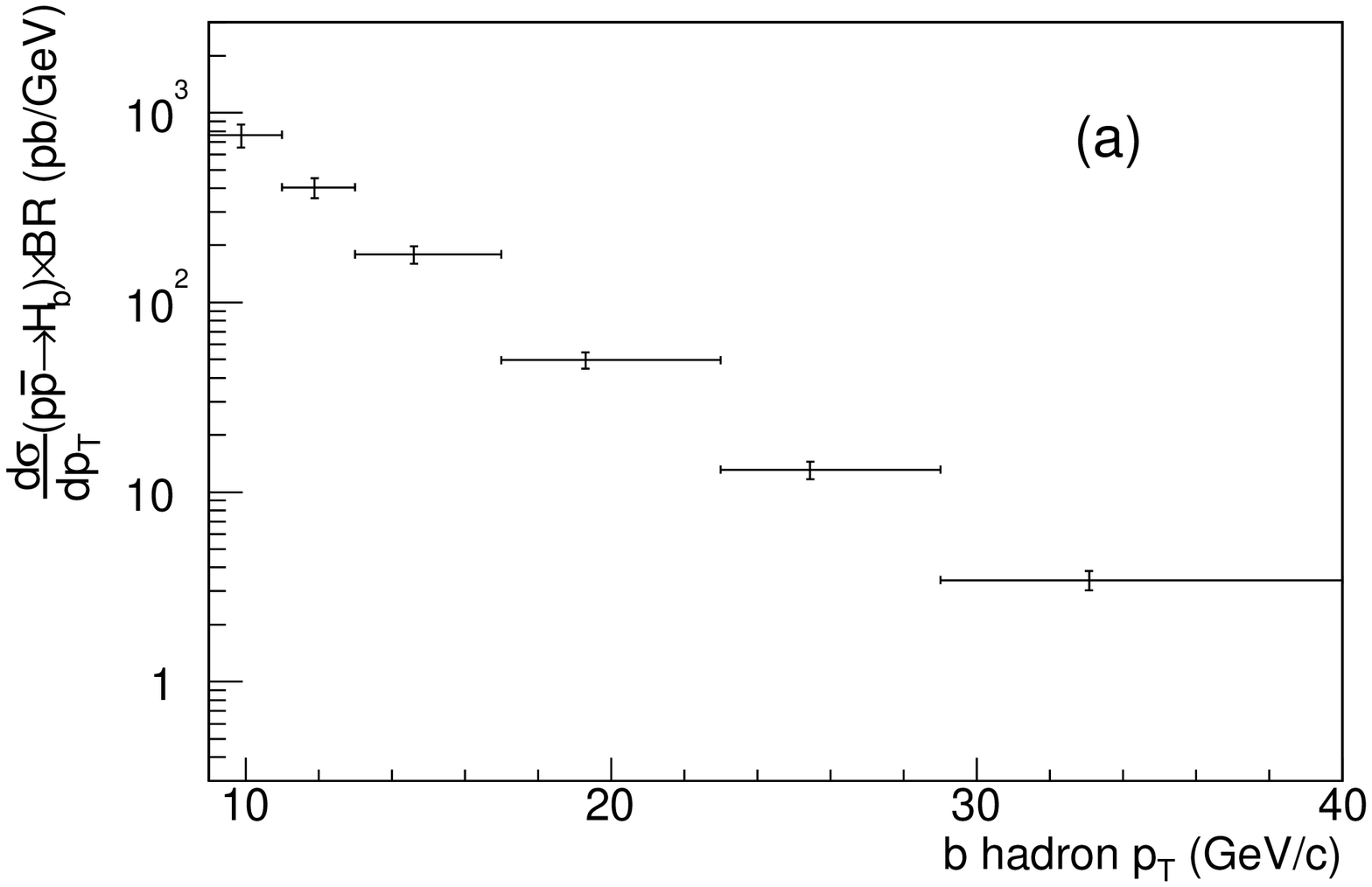,width=0.48\textwidth}
\psfig{figure=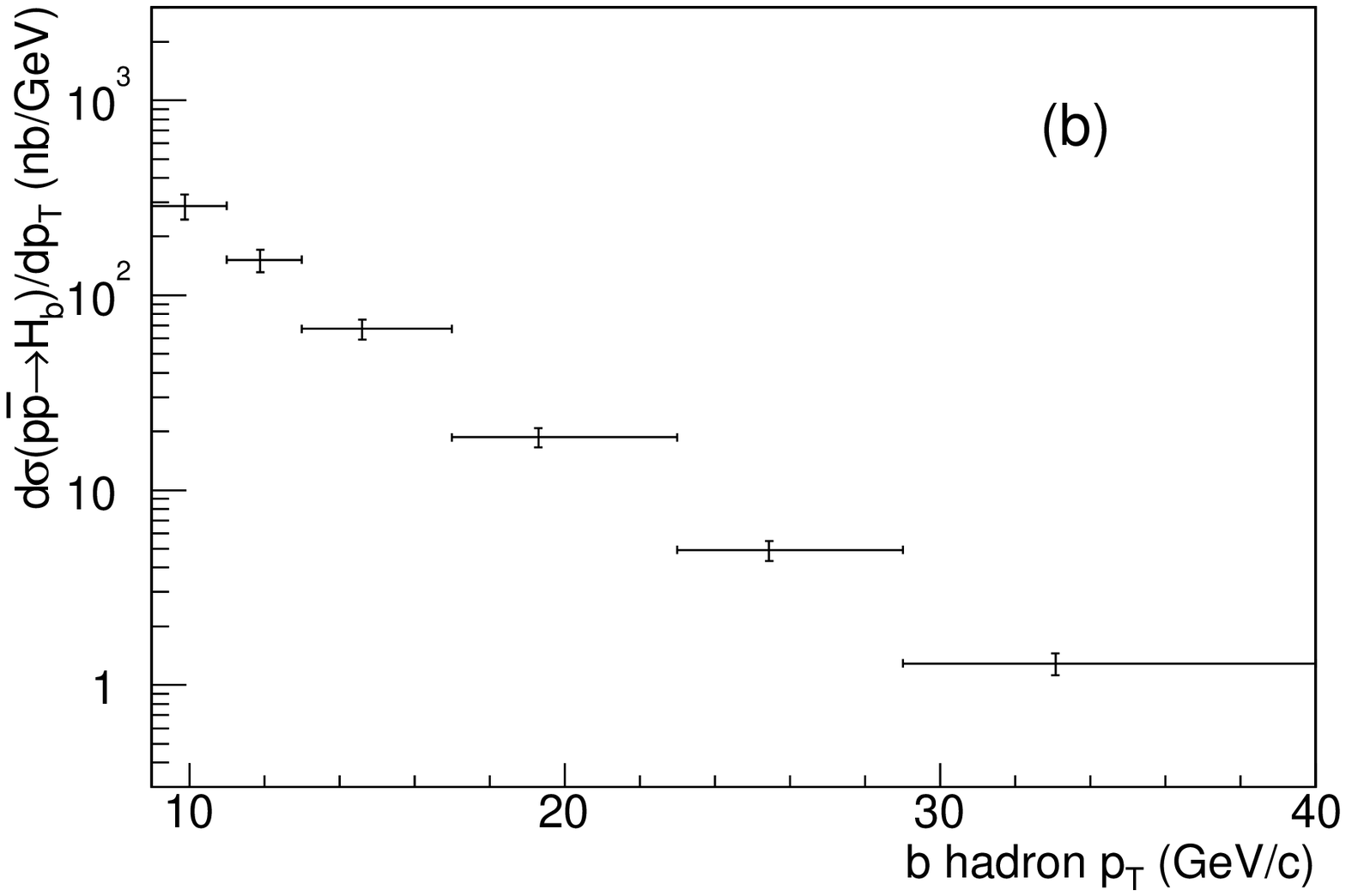,width=0.48\textwidth}
\caption{The differential cross section times branching ratio for $H_b \to \mu^- D^{0} X$, $D^0 \to K^- \pi^+$ is shown on the left,  
where 
$BR$ is shorthand notation for the product of branching ratios, 
$BR = \mathcal{B}(H_b\rightarrow \mu^- D^0 X)\times 
\mathcal{B}(D^0\rightarrow K^-\pi^+).$  The uncertainties shown on each point
include statistical and systematic uncertainties combined in quadrature.
Incorporating measured branching 
ratios \cite{PDG}, the differential cross section is shown on the right. The
uncertainties on each point include statistical, systematic and branching 
ratio uncertainties added in quadrature. 
\label{diffcorr}}

\end{figure}

After applying all corrections to the data, we obtain the total cross section times branching ratio 

\begin{displaymath}
\sigma(p\overline{p}\rightarrow H_b X)\times \mathcal{B}(H_b\rightarrow \mu^- D^0 X)\times \mathcal{B}(D^0\rightarrow K^-\pi^+)= 3.46  \pm 0.14 {\rm (stat)} ^{+0.36}_{-0.38} {\rm (syst)}\, {\rm nb}
\end{displaymath}
for $b$ hadrons with $p_T > 9\, \mbox{GeV}/c$ and $|y|<0.6$.  Using the
current world average  
branching ratios $\mathcal{B}(H_b \to \mu^- D^{0})=(6.84\pm 0.35)\%$ \cite{delphi,PDG} 
and  
$\mathcal{B}(D^0 \to K^- \pi^+)=(3.89\pm 0.05) \%$ \cite{D0Br, PDG}, we measure a 
total cross section of 

\begin{displaymath}
\sigma(p\overline{p}\rightarrow H_b X)= 1.30  \pm 0.05 {\rm (stat)} \pm 0.14  {\rm (syst)} 
\pm 0.07  (\mathcal{B})\,  \mu {\rm b}
\end{displaymath}
for $b$ hadrons with $p_T > 9\,  \mbox{GeV}/c$ 
and $|y|<0.6$.  The differential cross section times branching ratio is shown in 
Table~\ref{XsecTable}, 
and displayed in Fig.~\ref{diffcorr}(a).  Figure~\ref{diffcorr}(b) 
shows   
the differential cross section.  
 
\section{$H_b \to \mu^- {D^*}^+X$ Cross Section \label{dstarcrossSectionStuff}}

For the $H_b \to \mu^- {D^*}^+$, 
$ {D^*}^+ \to D^0 \pi^+$, $D^0 \to K^- \pi^+ $ channel,
the efficiencies are the same as those for the $H_b \to \mu^- D^{0}$, 
$D^0 \to K^-\pi^+$, except that we must additionally
account for the soft pion identification efficiency.

The systematics for the ${D^*}^+$ measurement are the 
same as those for the $D^0$ measurement listed in  
Table~\ref{SystematicsTable} unless mentioned specifically.

\begin{figure}
\psfig{figure=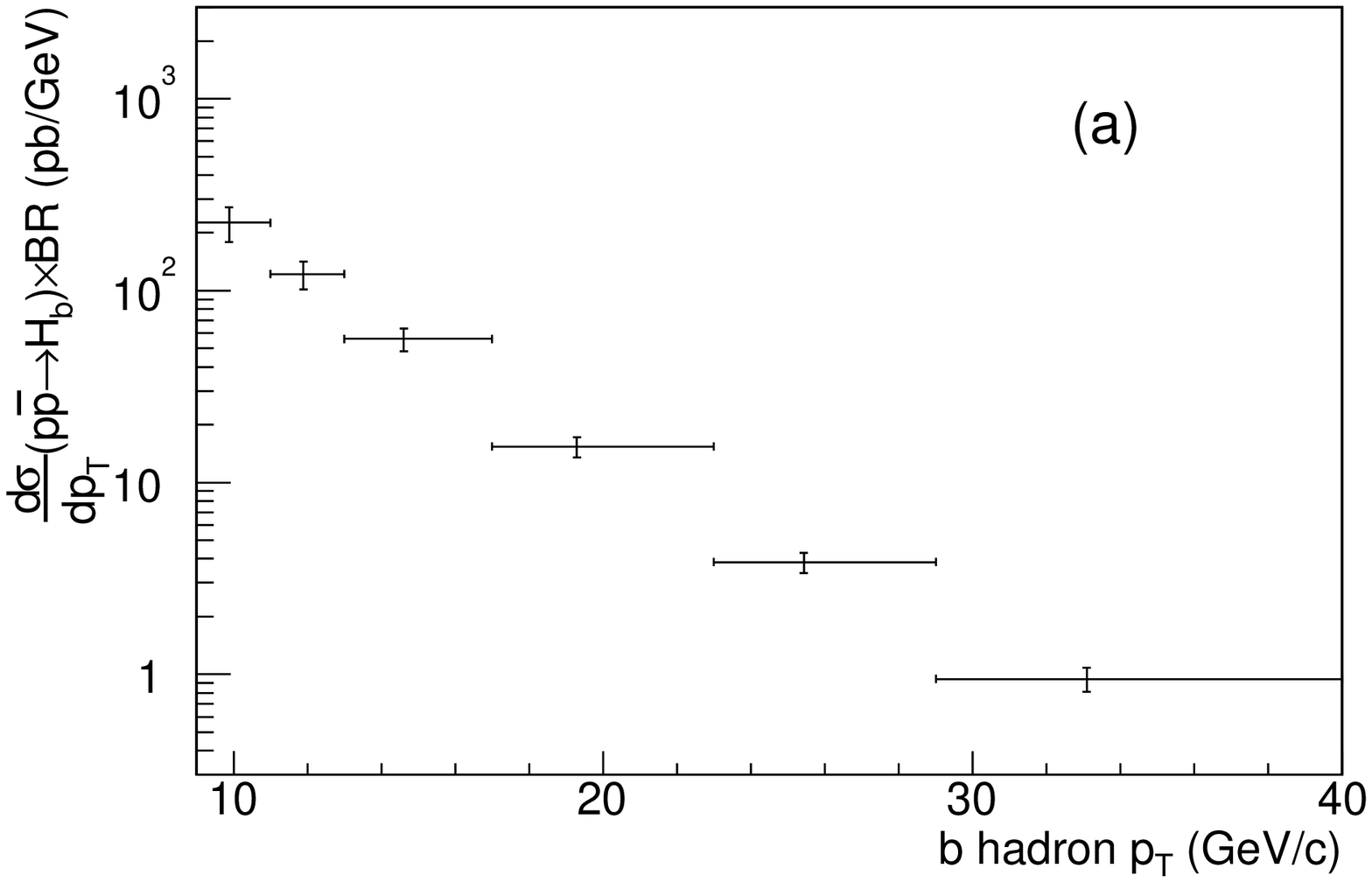,width=0.48\textwidth}
\psfig{figure=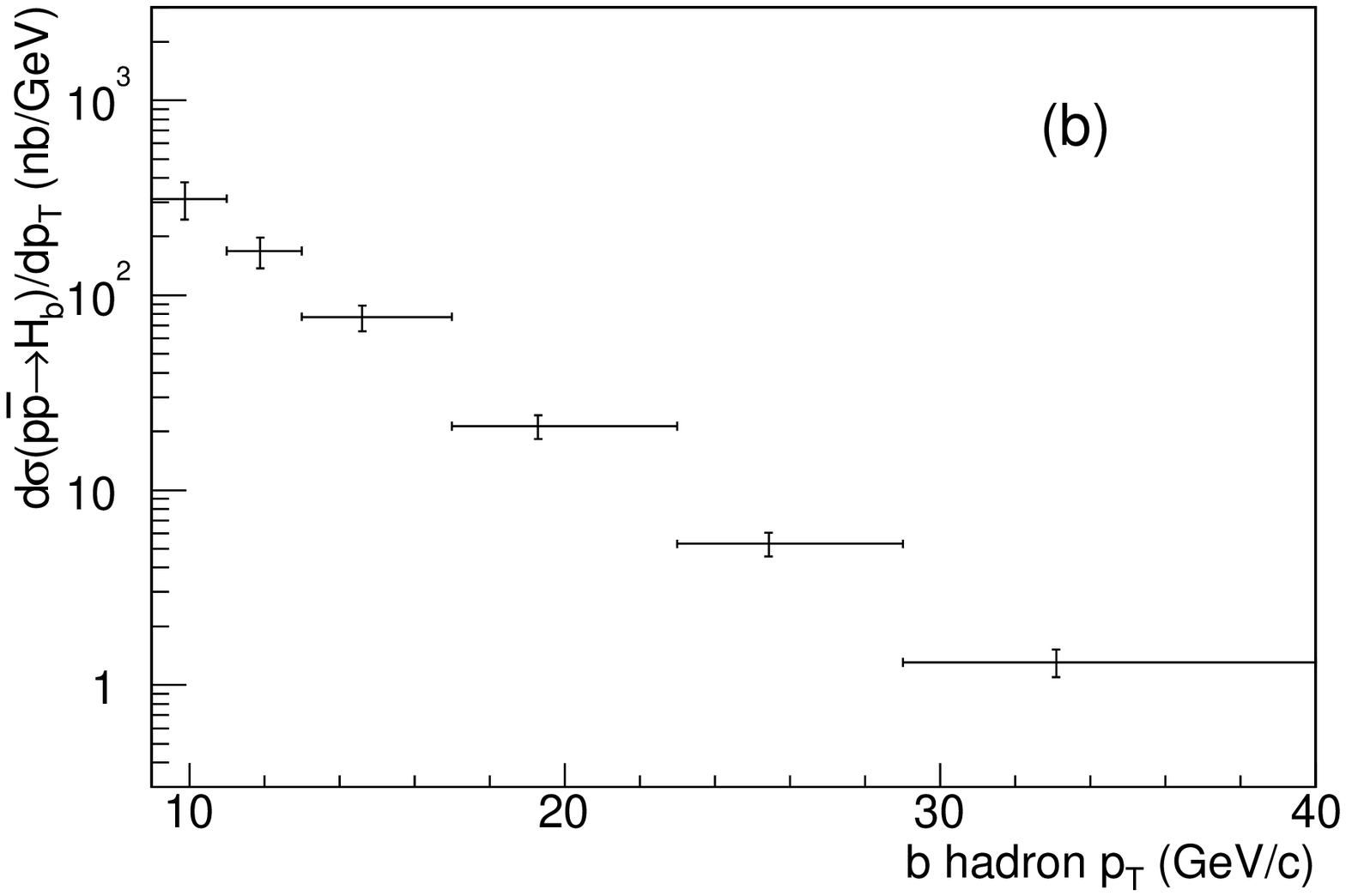,width=0.48\textwidth}
\caption{The differential cross section times branching ratio for 
$H_b \to \mu^- {D^*}^+ X$, ${D^*}^+ \to D^0 \pi_{soft}^+$, $D^0 \to K^- \pi^+$ 
is shown on the left,  
where $BR$ is shorthand notation for the product of branching ratios, 
$BR = \mathcal{B}(H_b\rightarrow \mu^- D^0 X)\times 
\mathcal{B}(D^0\rightarrow K^-\pi^+)\times \mathcal{B}({D^*}^+\rightarrow D^0 \pi^+).$  The uncertainties shown on each point
include statistical and systematic uncertainties combined in quadrature.
Incorporating measured branching ratios \cite{PDG},
the differential cross section is shown on the right. The
uncertainties on each point include statistical, systematic and branching 
ratio uncertainties added in quadrature. 
\label{difftimesBrDstar}}
\end{figure}

After applying all corrections to the data, we obtain the total cross section times branching ratio 
$H_b \to \mu^- {D^*}^+$, with $ {D^*}^+ \to D^0 \pi^+$ and $D^0 \to K^-\pi^+ $ decays  of

\begin{displaymath}
\sigma(p\overline{p}\rightarrow H_b X)\times \mathcal{B}(H_b \to \mu^- K^-\pi^+\pi^+ X) = 1.05  \pm 0.08 {\rm (stat)} ^{+0.13}_{-0.15} {\rm (syst)}\  {\rm nb},
\end{displaymath}
for $b$ hadrons with $p_T > 9\,  \mbox{GeV}/c$ and $|y|<0.6$ and using the
notation $\mathcal{B}(H_b \to \mu^- K^-\pi^+\pi^+ X)$ as shorthand 
for the product of 
$\mathcal{B}(H_b \to \mu^- {D^*}^+)$, 
$\mathcal{B}({D^*}^+ \to D^0 \pi^+)$, and 
$\mathcal{B}(D^0 \to K^-\pi^+)$.  Using the 
current world average  
branching ratios $\mathcal{B}(H_b \to \mu^- {D^*}^+)=(2.75\pm 0.19)\% $ \cite{delphi,PDG}, 
$\mathcal{B}( {D^*}^+ \to D^0 \pi^+)=(67.7\pm 0.5)\% $ \cite{DsBr, PDG}, and $\mathcal{B}(D^0 \to K^-\pi^+)=
(3.89\pm 0.05)\%$ \cite{D0Br, PDG}, we find a 
total cross section of 

\begin{displaymath}
\sigma(p\overline{p}\rightarrow H_b X)= 1.45 \pm 0.11  {\rm (stat)} ^{+0.18}_{-0.21} {\rm (syst)} \pm 0.10 (\mathcal{B})\, \mu{\rm b}
\end{displaymath}
for $b$ hadrons with $p_T > 9\, \mbox{GeV}/c$ 
and $|y|<0.6$.  The differential cross section times branching ratio is shown in 
Table~\ref{XsecTableDstar}, 
and displayed in Fig.~\ref{difftimesBrDstar}.  We also show the 
differential cross section 
using branching ratios measured elsewhere.

We consider the $H_b\rightarrow \mu^-{D^*}^+X$ cross-section results
to be a modest extension and cross check of the primary 
$H_b\rightarrow \mu^- D^0X$ cross section result presented in
this paper.  The $\mu^- {D^*}^+ X$ sample 
is of limited statistical
power and, being a subset of
the $\mu^- D^0 X$ sample, there is nothing to be gained
by averaging the two results.

\section{Comparison with theory and previous measurements}
\begin{figure}
\psfig{figure=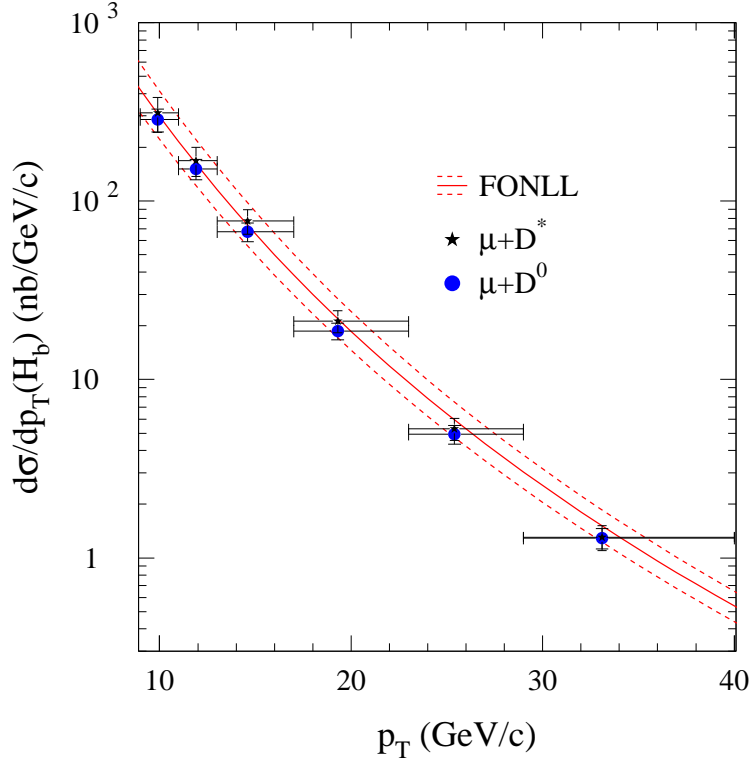,width=0.7\textwidth}
\caption{The $b$-hadron differential cross section for $|y(H_b)|<0.6$ from FONLL theory \cite{FONLL2} compared with measurements from $H_b \to \mu^- D^{0} X$ ($D^0 \to K^- \pi^+$) and  $H_b \to \mu^- {D^*}^+ X$ 
(${D^*}^+ \to D^0 \pi_{soft}^+$ and $D^0 \to K^- \pi^+$).
 \label{comboplot1}}
\end{figure}

Our results agree well with previous measurements
          of the $H_b$ cross section at CDF \cite{AcostaJpsi2005} 
and with the predictions of FONLL \cite{FONLL2}.  The 
CDF collaboration previously measured the  
$H_b$ cross 
section using $H_b \to J/\psi X$ 
of $1.34  \pm 0.07 \mu{\rm b}$ in the same range of $p_T$ 
and $y$ \cite{AcostaJpsi2005}.  The FONLL prediction for 
the total cross section with $p_T(b) > 9.0\, {\rm GeV}/c$ and $|y|<0.6$ is 
$1.39 ^{+0.49}_{-0.34}\, \mu {\rm b}$ \cite{FONLL2}.
Figure~\ref{comboplot1} shows the differential cross sections plotted together.   

\begin{figure}
\psfig{figure=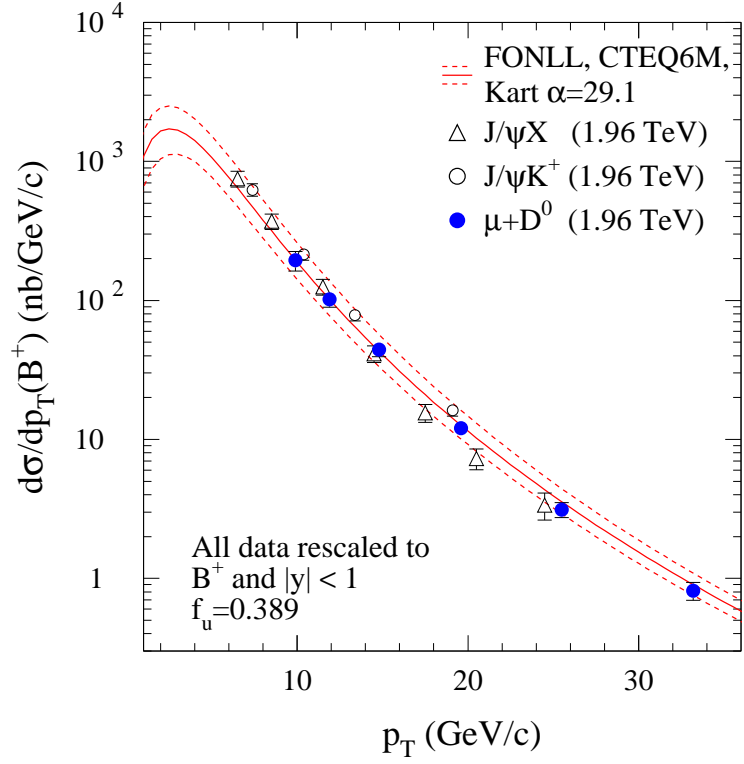,width=0.7\textwidth}
\caption{Comparison of differential $b$ cross-section results.  
The results presented are $H_b \rightarrow 
J/\psi X$ at $\sqrt{s}=1.96\, \rm  TeV$ \cite{AcostaJpsi2005},
$B^+ \rightarrow J/\psi K^+$ at $\sqrt{s}=1.96\, \rm TeV$ \cite{run2jpsik},
and
this result for $H_b\rightarrow \mu^- D^0 X$ at $\sqrt{s}=1.96\, \rm TeV$.
All results are scaled to $|\eta|<1$.  For comparison, the 
FONLL \cite{FONLL2} calculation
is included. \label{comboplot2}}
\end{figure}

To compare these results with the cross section measured in the
exclusive decay channel $B^+ \to J/\psi K^+$ \cite{run2jpsik}, we 
must account for the different rapidity ranges.  Reference~\cite{run2jpsik}
measured the cross section for $|y|<1$, while the result presented here
and the inclusive $J/\psi X$ result from Ref.~\cite{AcostaJpsi2005} 
measure the cross section for $|y|<0.6$.  Figure~\ref{comboplot2} 
shows the comparison between these results, where we have scaled all
cross sections to $|y|<1$ assuming that inclusive $H_b$ production 
is flat in rapidity.  Furthermore, we 
correct for the differences in fragmentation fractions for 
$B^+$ ($f_u$) and $H_b$ ($f_u+f_d+f_s$) \cite{annovi}.  
Figure~\ref{comboplot2}
shows good agreement between these recent $b$-hadron cross 
section measurements at $\sqrt{s}=1.96\, \rm TeV$, as well as 
agreement between these complementary measurements and 
the FONLL prediction.

The results presented here also compare favorably to the recent theoretical 
work  using the general-mass variable-flavor-number scheme (GM-VFNS) 
presented in Ref.~\cite{gmvfns}.

The trigger and analysis strategy shown here for the $\mu^- D^0$
and $\mu^- {D^*}^+$ final state are quite different than those
utilized in measurements using $J/\psi$ final states.  In addition
to being statistically independent, 
many of the systematic uncertainties of this result  
are distinct from those of  previous measurements.  This result therefore
provides a unique and independent measurement of the $H_b$ cross
section at the Tevatron.

\section{Conclusion}
\label{sec:conc}

We have measured the $b$-hadron production 
cross section in $H_b \to \mu^- D^0 X$ and 
$H_b \to \mu^- {D^*}^+ X$ decays using
the CDF II detector at the Fermilab Tevatron.   The muon plus 
charm hadron decay
signature was acquired using a dedicated trigger that takes advantage 
of the muon signature as well as the long $H_b$ lifetime.  By 
selecting an event sample based upon decay length 
information measured in the trigger, it is possible to 
acquire clean $H_b$ signal samples at lower $p_T(H_b)$ than have been
measured before.   Complementary data samples are used to measure 
the trigger efficiencies for this analysis and Monte Carlo simulation is 
used to measure the acceptance.  After accounting for backgrounds that
mostly arise from heavy flavor ($b\overline{b}$ and $c\overline{c}$ production)
the cross section is measured to be

\begin{displaymath}
\sigma(p\overline{p}\rightarrow H_b X)= 1.30  \pm 0.05 {\rm (stat)} \pm 0.14 {\rm (syst)} 
\pm 0.07  (\mathcal{B})\,  \mu {\rm b}
\end{displaymath}
for $b$ hadrons with $p_T > 9\, \mbox{GeV}/c$ 
and $|y|<0.6$.  
We additionally present the differential cross section
$d\sigma/dp_T$ in this kinematic region, which extends to lower
$p_T(H_b)$ than previous measurements in semileptonic modes.  
We find the results presented here to be in good agreement
with other recent measurements of the $H_b$ cross section and in good
agreement with fixed order next-to-leading logarithm calculations.

\section{Acknowledgements}
 
We thank the Fermilab staff and the technical staffs of the participating institutions for their vital contributions. This work was supported by the U.S. Department of Energy and National Science Foundation; the Italian Istituto Nazionale di Fisica Nucleare; the Ministry of Education, Culture, Sports, Science and Technology of Japan; the Natural Sciences and Engineering Research Council of Canada; the National Science Council of the Republic of China; the Swiss National Science Foundation; the A.P. Sloan Foundation; the Bundesministerium f\"ur Bildung und Forschung, Germany; the Korean Science and Engineering Foundation and the Korean Research Foundation; the Science and Technology Facilities Council and the Royal Society, UK; the Institut National de Physique Nucleaire et Physique des Particules/CNRS; the Russian Foundation for Basic Research; the Ministerio de Ciencia e Innovaci\'{o}n, and Programa Consolider-Ingenio 2010, Spain; the Slovak R\&D Agency; and the Academy of Finland.

\pagebreak

\begin{table}
\begin{center}
\caption{The signal yields for the number of $D^0$ and ${D^*}^+$ events per 
$p_T(\mu^- D^0)$ bin before (and after) correcting for SVT efficiency.   
Since the SVT efficiency depends upon the relative track isolation,
a correction is performed based upon the topology of the signal events.
\label{yieldtable}}
\begin{tabular}{rrrrr}
\hline \hline
$p_T(\mu^- D^0)$ [GeV/$c$] & \multicolumn{2}{c}{$D^0$ signal yield} & \multicolumn{2}{c}{$D^{*}$ signal yield} \\
 & Measured & Corrected & Measured  & Corrected \\
\hline
$9$-$11$  & 867.9 $\pm$ 53.5 & 1040.6 $\pm$ 64.1 & 82.3 $\pm$ 10.5  &\  \  96.1 $\pm$ 12.2  \\
$11$-$13$  & 863.1 $\pm$ 45.8 & 1034.4 $\pm$ 54.9 &142.6 $\pm$ 13.2  & 170.0 $\pm$ 15.7\\
$13$-$17$  &\ \ \  1016.8 $\pm$ 46.3 & \ \ \ 1192.4 $\pm$ 54.3 &\ \ \ 236.4 $\pm$ 16.9 & \ \ \ 283.4 $\pm$ 20.2\\
$17$-$29$  & 669.6 $\pm$ 38.3 & 781.9 $\pm$ 44.7 & 169.6 $\pm$ 14.1 & 199.8 $\pm$ 16.6\\
$29$-$40$  & 67.0\ $\pm$ 12.2  & 80.1 $\pm$ 14.6 & 14.3  $\pm$ \ 4.2  & 18.7 $\pm$ \ 5.6  \\
\hline \hline
\end{tabular}
\end{center}
\end{table}

\begin{table}[!hpt]
\begin{center}
\singlespacing
\caption{The acceptances times efficiencies using 
the CDF phenomenological spectrum (CDF Sample) and MRSD$_0$ NLO (MRSD$_0$ Sample) 
MC simulation. The average
$p_T(H_b)$ for the two spectra vary by no more than $1\% $ in any transverse momentum bin. \label{acctimeseff}}
\begin{tabular}{rccc}
\hline \hline
$p_T(H_b)$ [GeV/$c$] &\ \ \ \ Average $p_T(H_b)$ [GeV/$c$]\ \ \ \ \  & \multicolumn{2}{c}{Overall efficiency} \\
 & & CDF Sample & MRSD$_0$ Sample\\
\hline
$9$-$ 11$ & \ 9.9  & \  \   $1.47\times 10^{-3}$ & \  \  $ 1.51\times 10^{-3}$ \\
$11$-$ 13$& 11.9 & \  \ $6.12\times 10^{-3}$ & \  \  $ 6.16\times 10^{-3}$ \\
$13$-$ 17$& 14.8 & \  \ $1.38\times 10^{-2}$ & \  \ $1.40 \times 10^{-2}$\\
$17$-$ 23$& 19.6 & \  \ $2.68\times 10^{-2}$ & \  \ $2.70\times 10^{-2}$ \\
$23$-$ 29$& 25.5 & \  \ $3.90\times 10^{-2}$ & \  \ $3.84\times 10^{-2}$ \\
$29$-$ 40$& 33.2 & \  \ $4.61\times 10^{-2}$ & \  \ $4.63\times 10^{-2}$ \\
\hline \hline
\end{tabular}
\end{center}

\end{table}


\begin{table}[!hpt]
\begin{center}
\caption{The weights $w_{ij}$ from the CDF phenomenological spectrum 
MC simulation with all momenta in GeV/$c$.\label{CDFweights}}
\begin{tabular}{crrrrr}
\hline \hline 
$w_{ij}$ & \multicolumn{5}{c}{$p_T(\mu^- D^0)$} \\ 
 &$9$-$11$\ \ \ \ \  & $11$-$13$\ \ \ \ \ & $13$-$17$\ \ \ \ \ &
$17$-$29$\ \ \ \ \ &$29$-$40$\ \ \ \ \  \\
\hline
$H_b \to  DD$, $H_b \to  D \tau$&$\ \ \ 5.54\times 10^{-2}$ &\ \ \  $5.81\times 10^{-2}$ & 
\ \ \ $5.94 \times 10^{-2}$&\ \ \  $5.64\times 10^{-2}$ &\ \ \  $1.84\times 10^{-2}$ \\
$p_T(H_b)<9$     &$3.26\times 10^{-3}$ & $<1\times 10^{-6}$ & $<1\times 10^{-6}$ & $<1\times 10^{-6}$ &$<1\times 10^{-6}$  \\
$9$-$ 11$   &$3.02\times 10^{-1}$ & $1.34\times 10^{-3}$ 
& $<1\times 10^{-6}$& $<1\times 10^{-6}$ &$<1\times 10^{-6}$ \\
$11$-$ 13$  &$3.78\times 10^{-1}$ & $2.65\times 10^{-1}$ 
& $1.00\times 10^{-4}$ & $<1\times 10^{-6}$ & $<1\times 10^{-6}$ \\
$13$-$ 17$  &$2.22\times 10^{-1}$ & $5.27\times 10^{-1}$ 
& $4.32\times 10^{-1}$ & $5.77\times 10^{-4}$ & $<1\times 10^{-6}$ \\
$17$-$ 23$  &$3.70\times 10^{-2}$ & $1.31\times 10^{-1}$ 
& $4.35\times 10^{-1}$ & $3.73\times 10^{-1}$ & $<1\times 10^{-6}$ \\
$23$-$ 29$ &$3.39\times 10^{-3}$ & $1.51\times 10^{-2}$ 
& $6.25\times 10^{-2}$ & $3.59\times 10^{-1}$ & $1.69\times 10^{-3}$ \\
$29$-$ 40$  &$6.00\times 10^{-5}$ & $1.51\times 10^{-3}$ 
& $1.04\times 10^{-2}$ & $1.93\times 10^{-1}$ & $6.35\times 10^{-1}$ \\
$p_T(H_b)> 40$    &$<1\times 10^{-6}$ & $1.21\times 10^{-4}$ 
& $3.83\times 10^{-4}$ & $1.78\times 10^{-2}$ & $3.45\times 10^{-1}$ \\

\hline \hline
\end{tabular}
\end{center}
\singlespacing

\end{table}

\begin{table}[!hpt]
\begin{center}
\caption{The weights $w_{ij}$ from the NDE+MRSD$_0$ MC simulation with all momenta in GeV/$c$.\label{MRSD0weights}}
\begin{tabular}{crrrrr}
\hline \hline
$w_{ij}$ & \multicolumn{5}{c}{$p_T(\mu^- D^0)$ [GeV/$c$]} \\ 
 &$9$-$11$\ \ \ \ \  & $11$-$13$\ \ \ \ \ & $13$-$17$\ \ \ \ \ &
$17$-$29$\ \ \ \ \ &$29$-$40$\ \ \ \ \  \\
\hline
$H_b \to  DD$, $H_b \to D \tau$ & \ \ $6.09\times 10^{-2}$ & \ \ $6.26\times 10^{-2}$ & \ \
$6.39 \times 10^{-2}$& \ \ $6.50\times 10^{-2}$ & \ \  
$3.14\times 10^{-2}$ \\
$p_T(H_b)<9\, {\rm GeV}/c$      &$2.06\times 10^{-3}$ & $<1\times 10^{-6}$ & $<1\times 10^{-6}$ & $<1\times 10^{-6}$ &$<1\times 10^{-6}$  \\
$9$-$ 11\, {\rm GeV}/c$  &$2.75\times 10^{-1}$ & $1.38\times 10^{-3}$ 
& $<1\times 10^{-6}$& $<1\times 10^{-6}$ &$<1\times 10^{-6}$ \\
$11$-$ 13\, {\rm GeV}/c$ &$3.66\times 10^{-1}$ & $2.42\times 10^{-1}$ 
& $6.19\times 10^{-4}$ & $<1\times 10^{-6}$ & $<1\times 10^{-6}$ \\
$13$-$ 17\, {\rm GeV}/c$ &$2.43\times 10^{-1}$ & $5.33\times 10^{-1}$ 
& $3.88\times 10^{-1}$ & $9.40\times 10^{-5}$ & $<1\times 10^{-6}$ \\
$17$-$ 23\, {\rm GeV}/c$ &$4.98\times 10^{-2}$ & $1.42\times 10^{-1}$ 
& $4.66\times 10^{-1}$ & $3.65\times 10^{-1}$ & $<1\times 10^{-6}$ \\
$23$-$ 29\, {\rm GeV}/c$ &$3.29\times 10^{-3}$ & $1.63\times 10^{-2}$ 
& $6.82\times 10^{-2}$ & $3.61\times 10^{-1}$ & $1.21\times 10^{-3}$ \\
$29$-$ 40\, {\rm GeV}/c$ &$4.71\times 10^{-4}$ & $2.28\times 10^{-3}$ 
& $1.27\times 10^{-2}$ & $1.91\times 10^{-1}$ & $6.00\times 10^{-1}$ \\
$p_T(H_b)> 40\, {\rm GeV}/c$    &$<1\times 10^{-6}$ & $7.90\times 10^{-5}$ 
& $7.30\times 10^{-4}$ & $1.76\times 10^{-2}$ & $3.67\times 10^{-1}$ \\
\hline \hline
\end{tabular}
\end{center}

\end{table}

\begin{table}[!hpt]
\begin{center}
\caption{The systematic uncertainties of the $d\sigma(p\overline{p} \to H_b)/dp_{T}$ measurement.  All numbers listed in percent.  \label{SystematicsTable}}
\renewcommand{\arraystretch}{1.25}
\begin{tabular}{lccccccc}

\hline \hline
 &  \multicolumn{7}{c}{$p_T$ range [GeV/$c$]} \\

Source & All &\ \    $9$-$11$\ \   & \ \   $11$-$13$\ \   & \ \  $13$-$17$\ \   & 
 \ \  $17$-$23$ \ \  & \ \  $23$-$29$ \ \  & \ \  $29$-$40$ \ \  \\
\hline
Luminosity       & 6.0               & 6.0        & 6.0         & 6.0        & 6.0        & 6.0        & 6.0 \\
\hline
L1 Efficiency    & 2.6               & 2.6        & 2.6         & 2.6        & 2.5        & 2.5        & 2.5 \\
L2 XFT Efficiency& 1.0               & 1.0        & 1.0         & 1.0        & 1.0        & 1.0        & 1.0 \\
L2 SVT Efficiency& 1.3               & 1.6        & 1.2         & 0.8        & 0.7        & 0.7        & 0.9 \\
L3 Efficiency    & 0.2               & 0.2        & 0.2         & 0.2        & 0.2        & 0.1        & 0.1 \\
\hline
CMU Efficiency   & 1.6               & 1.6        & 1.6         & 1.6        & 1.6        & 1.6        & 1.6 \\
CMP Efficiency   & $^{+1.4}_{-1.3}$ & $^{+1.4}_{-1.3}$ & $^{+1.4}_{-1.3}$ & $^{+1.4}_{-1.3}$ & 1.4 & 1.5 & 1.7 \\
SVX~II Efficiency   & $^{+1.5}_{-1.0}$ & $^{+1.5}_{-1.0}$ &$^{+1.5}_{-1.0}$ &$^{+1.5}_{-1.0}$ &$^{+1.5}_{-1.1}$ &$^{+1.5}_{-1.1}$ &$^{+1.5}_{-1.0}$ \\
SVX~II 2nd track Efficiency & 1.0               & 1.0        & 1.0         & 1.0        & 1.0        & 1.0        & 1.0 \\
COT Eff. ($D^0$)  & $^{+3.1}_{-4.1}$ & $^{+3.2}_{-4.3}$ & $^{+3.3}_{-4.2}$ & $^{+3.0}_{-3.8}$ & $^{+2.9}_{-3.5}$ & $^{+2.8}_{-3.3}$ & $^{+2.7}_{-3.1}$ \\
COT Eff. ($\mu^- {D^*}^+$) & $^{+7.0}_{-8.9}$ &$^{+7.9}_{-9.7}$ &$^{+7.1}_{-9.6}$ &$^{+6.1}_{-8.2}$ &$^{+5.3}_{-6.1}$ &$^{+4.9}_{-5.0}$ &$^{+4.5}_{-4.3}$ \\
\hline
Vertex + Fit Eff & 0.4               & 0.4        & 0.4         & 0.4        & 0.4        & 0.4        & 0.4 \\
$c\overline{c}$ Background & 4.7     & 5.3        & 5.8         & 3.8        & 2.5        & 2.1        & 1.7 \\
$b\overline{b}$ Background & 0.6     & 0.6        & 0.6         & 0.6        & 0.6        & 0.6        & 0.6 \\

$d_0$ Smearing   & $^{+0.9}_{-2.6}$ & $^{+1.1}_{-2.6}$ & $^{+0.8}_{-2.8}$ & $^{+0.7}_{-2.6}$ & $^{+0.8}_{-2.3}$ & $^{+1.0}_{-2.5}$ & $^{+1.5}_{-1.4}$ \\
final state radiation           & 1.2 & 1.0 & 1.8 & 1.8 & 0.7 & 2.2 & 4.0 \\
$\mathcal{B}(H_b \to \mu^- D^{**} X)$  $(\mu^- D^0)$    & 1.9 & 2.1 & 1.7 & 1.6 & 2.0 & 1.5 & 1.7 \\
$\mathcal{B}(H_b \to \mu^- D^{**} X)$ $(\mu^- {D^*}^+)$  & 1.9 & 2.0 & 1.9 & 1.8 & 2.3 & 1.7 & 2.0 \\
$\mathcal{B}(H_b \to DD)$, $\mathcal{B}(H_b \to D\tau)$       & 2.4 & 2.4 &2.4  & 2.5 & 2.6 & 2.3 & 2.4 \\
MC stat. ($\mu^- D^0$)          & 0.5 & 1.7 & 1.1 & 0.9 & 1.0 & 1.6 & 2.1 \\
MC stat. ($\mu^- {D^*}^+$)        & 1.2 & 5.8 & 3.4 & 2.2 & 2.2 & 3.3 & 4.4 \\
MC $p_T$ shape ($\mu^- D^0$)    & 3.4 & 5.7 & 3.0 & 1.7 & 2.2 & 2.0 & 2.5 \\
MC $p_T$ shape ($\mu^- {D^*}^+$) & 4.4 & 5.7 &4.4 &3.9 &4.0 &0.6 &2.8 \\
\hline 
Total ($\mu^- D^0$) &$^{+10.4}_{-11.0}$  &$^{+11.9}_{-12.4}$ & $^{+11.0}_{-11.5}$ & $^{+9.6}_{-10.1}$ &$^{+9.2}_{-9.6}$ &$^{+9.2}_{-9.6}$ &$^{+10.0}_{-10.1}$ \\  \hline

Total ($\mu^- {D^*}^+$) & $^{+12.5}_{-13.9}$ &$^{+14.9}_{-16.0}$ & $^{+13.5}_{-15.1}$ &$^{+11.7}_{-13.1}$ &$^{+11.0}_{-11.5}$ & $^{+10.3}_{-10.5}$ &$^{+11.5}_{-11.3}$ \\

\hline \hline
\end{tabular}
\end{center}
\singlespacing

\end{table}

\begin{table}[!hpt]
\begin{center}
\caption{Differential cross section $d\sigma(p\overline{p} \to H_b)/dp_{T}$ 
and differential cross section times branching fraction of $H_b \to \mu^- D^0 X$, $D^0 \to K^- \pi^+$, where the first uncertainty is statistical, and 
the second is systematic. In the case of the differential
cross section, the third uncertainty arises from the applied
branching fractions.\label{XsecTable}}
\renewcommand{\arraystretch}{1.25}
\begin{tabular}{cll}
\hline \hline
$p_T$ bin [GeV/$c$] &\ \ \ \ \  $d \sigma/dp_{T} \times \mathcal{B}$ [pb/(GeV/$c$)]  & \ \ \ \  $d\sigma/dp_T$  [pb/(GeV/$c$)]\\
\hline
9-11 & \ \ \ \ \ \  $762 \pm 50 ^{+91}_{-94}$ &\ \ \ \  $287\pm 19 ^{+34}_{-35} \pm 15$\\
11-13 & \ \ \ \ \ \ $403 \pm 18 ^{+44}_{-44}$ &\ \ \ \   $152\pm 7 \pm 17 \pm 8$\\
13-17 & \ \ \ \ \ \ $179 \pm  6 ^{+17}_{-18}$ &\ \ \ \   $67.3\pm 2.3^{+6.4}_{-6.8} \pm 3.6$\\
17-23 & \ \ \ \ \ \ $49.7 \pm 1.5 ^{+4.6}_{-4.8}$ &\ \ \ \   $17.8\pm 0.6 ^{+1.7}_{-1.8} \pm 0.9$\\
23-29 & \ \ \ \ \ \ $13.1 \pm 0.6 ^{+1.2}_{-1.3}$ &\ \ \ \   $4.93\pm 0.23 ^{+0.45}_{-0.49}\pm 0.26$\\
29-40 & \ \ \ \ \ \  $3.43 \pm 0.20 ^{+0.34}_{-0.35}$ &\ \ \ \   $1.29\pm 0.08\pm 0.13\pm 0.07$\\

\hline \hline
\end{tabular}
\end{center}

\end{table}

\begin{table}[!hpt]
\begin{center}
\caption{Differential cross section $d\sigma(p\overline{p} \to H_b)/dp_{T}$ and
differential cross section 
times branching fraction of
$H_b \to \mu^- {D^*}^+ X$, ${D^*}^+ \to D^0 \pi^+$, $D^0 \to K^- \pi^+$, 
where  
the first uncertainty is statistical, and the second is systematic.
 In the case of the differential
cross section, the third uncertainty arises from the applied
branching fractions.\label{XsecTableDstar}}
\renewcommand{\arraystretch}{1.25}
\begin{tabular}{cll}
\hline \hline
$p_T$ bin [GeV/$c$] & \ \ \ \ \  
$d \sigma/dp_{T} \times \mathcal{B}$ [pb/(GeV/$c$)]
& \ \ \ \  $d\sigma/dp_T$  [pb/(GeV/$c$)]\\
\hline
9-11 & \ \ \ \ \ \  $226 \pm 30 ^{+34}_{-36}$ 
&\ \ \ \ $312\pm 41 ^{+47}_{-50} \pm 22$ \\
11-13 & \ \ \ \ \ \  $122 \pm 10 ^{+16}_{-18}$ & 
\ \ \ \ $168 \pm 14 ^{+22}_{-25}\pm 12 $ \\
13-17 & \ \ \ \ \ \  $56.0 \pm 3.0^{+6.5}_{-7.3}$ & 
\ \ \ \  $77.3 \pm 4.1 ^{+9.0}_{-10.1}\pm 5.5 $\\
17-23 & \ \ \ \ \ \  $15.4 \pm 0.7 ^{+1.7}_{-1.8}$& 
\ \ \ \ $ 21.3 \pm 1.0 ^{+2.3}_{-2.5}\pm 1.5 $  \\
23-29 & \ \ \ \ \ \  $3.84 \pm 0.25 ^{+0.40}_{-0.40}$ & 
\ \ \ \ $ 5.30 \pm 0.35 \pm 0.55 \pm 0.37 $ \\
29-40 & \ \ \ \ \ \   $0.95 \pm  0.09 ^{+0.11}_{-0.11} $ & 
\ \ \ \ $ 1.31 \pm 0.12\pm 0.15 \pm 0.09 $ \\

\hline \hline
\end{tabular}
\end{center}

\end{table}

\newpage

\newpage

\newpage

\newpage

\end{document}